\documentclass[onecolumn,amsmath,tightenlines,amssymb,11pt,superscriptaddress,nofootinbib]{revtex4}

\usepackage{graphicx}
\usepackage{amsmath,amssymb,amsfonts,amsthm,stmaryrd,mathtools,bm,physics,tensor}
\usepackage{color}
\usepackage{tikz}
\usepackage[normalem]{ulem}
\allowdisplaybreaks[1]

\usepackage[bookmarks,linktocpage, colorlinks=true, plainpages = false, citecolor = treegreen,  linkcolor=darkblue, urlcolor = darkblue, filecolor = blue]{hyperref} 
\usepackage{orcidlink}

\usepackage[capitalise]{cleveref}
\crefformat{equation}{Eq.~(#2#1#3)}
\crefformat{table}{Tab.~(#2#1#3)}
\crefformat{figure}{Fig.~(#2#1#3)}
\crefformat{appendix}{App.~(#2#1#3)}
\crefformat{section}{Sec.~(#2#1#3)}
\crefformat{subsection}{Subsec.~(#2#1#3)}

\def\be#1\ee{\begin{align}#1\end{align}}

\def\ba{\begin{eqnarray}}
	\def\ea{\end{eqnarray}}
\def\nn{\nonumber}

\definecolor{tealgreen}{rgb}{0.0, 0.5, 1.0}
\definecolor{darkblue}{rgb}{0., 0.4, 0.8}
\definecolor{cadmiumred}{rgb}{1., 0., 0.22}
\definecolor{treegreen}{rgb}{0., 0.7, 0.3}

\begin{document}
	
\title{Spherically-symmetric solutions in quasi-local Einstein-Weyl gravity}

\author{Johanna Borissova\,\orcidlink{0000-0001-7125-4372}}
\email{j.borissova@imperial.ac.uk}
\affiliation{Abdus Salam Centre for Theoretical Physics, Imperial College London, London SW7 2AZ, UK}

\author{Breno L. Giacchini\,\orcidlink{0000-0002-6449-7189}}
\email{breno.giacchini@matfyz.cuni.cz}
\affiliation{Institute of Theoretical Physics, Faculty of Mathematics and Physics, Charles University, V Hole{\v s}ovi{\v c}k{\'a}ch 2, 180 00 Prague 8, Czech Republic}

\author{Aaron Held\,\orcidlink{0000-0003-2701-9361}}
\email{aaron.held@phys.ens.fr}
\affiliation{
Institut de Physique Théorique Philippe Meyer, Laboratoire de Physique de l’\'Ecole normale sup\'erieure (ENS), Universit\'e PSL, CNRS, Sorbonne Universit\'e, Universit\'e Paris Cité, F-75005 Paris, France
}

\begin{abstract}
{\sc Abstract:} 
Quantum-gravitational effective actions with higher-derivative and non-local operators are expected to regularize the singularities of general relativity. 
 Here we focus on quasi-local Einstein-Weyl gravity and obtain a local classification of static spherically symmetric solutions expandable as Frobenius series in Kundt coordinates.
In contrast to local Einstein-Weyl gravity, and more generally quadratic gravity, we find that the quasi-local theory admits only regular solutions at the radial core.
In addition, we find asymptotic $1/r^{6}$-corrections to the Schwarzschild geometry at large radial distances. 
Other solution classes around generic expansion points describe Schwarzschild-like and other types of horizons, as well as symmetric and non-symmetric wormhole throats. 
\end{abstract}

\maketitle
\tableofcontents

\bigskip

\bigskip

\section{Introduction}

The Einstein-Hilbert action of general relativity (GR) provides a remarkably complete description of gravitational dynamics~\cite{Penrose:1969pc,Wald:1997wa} --- at least outside of black-hole horizons and whenever quantum effects are negligible. 
Inside of horizons, however, the theory predicts singularities and thus its own breakdown~\cite{Penrose:1964wq,Hawking:1973uf}.
At the same time, close to such singularities, there is no reason to expect that quantum corrections  remain small.  One may expect the inclusion of quantum corrections in the gravitational effective action to regularize black-hole singularities, although addressing this question in a fully non-linear and global treatment of solutions has largely remained elusive. 
For overviews on the current status and challenges to black holes, both in classical and quantum gravity, see e.g.~\cite{Berti:2015itd,Barack:2018yly,Cardoso:2019rvt,Ayzenberg:2023hfw,Carballo-Rubio:2025fnc,Yunes:2025xwp,Buoninfante:2024oxl,Buoninfante:2024yth,Bambi:2025wjx}.
As reviewed below, quantum corrections are generally expected to comprise (i) local higher-curvature terms and (ii) additional non-local terms. The main motivation of this work is to explore the effect of non-local corrections on classical solutions in the reduced framework of static spherical symmetry. 
\\

From the perspective of perturbative quantum gravity, local higher-derivative terms are typically associated with renormalization (see, e.g.,~\cite{Buchbinder:2021wzv} for a pedagogical introduction).
For instance, higher-derivative gravity models can be (super)renormalizable~\cite{Stelle:1976gc,Asorey:1996hz}, whereas in the quantization of Einstein gravity one obtains an infinite tower of higher-derivative corrections, signalling that this theory is perturbatively non-renormalizable~\cite{tHooft:1974toh,Deser:1974cz,Deser:1974xq,Goroff:1985th}.
Nevertheless, quantum corrections to GR can be treated in the framework of effective field theory (EFT)~\cite{Donoghue:1994dn,Donoghue:2022eay}.
All of the above corrections are expected to manifest in a low-energy effective action which can be systematically organized as a double expansion in the Riemann curvature $R_{\mu\nu\rho\sigma}$ and/or external derivatives. This double expansion is often collapsed to an indiscriminate expansion in the dimensionality of operators with respect to a single cutoff scale $\ell$ in units of length. The effect of quadratic~\cite{Stelle:1977ry,Lu:2015cqa,Lu:2015psa,Kokkotas:2017zwt,Podolsky:2018pfe,Svarc:2018coe,Podolsky:2019gro,Held:2022abx,Giacchini:2025mlv}, cubic~\cite{deRham:2020ejn,Giacchini:2025gzw}, and quartic~\cite{Cardoso:2018ptl,Cano:2019ore,Cano:2020cao} corrections on stationary black-hole solutions has then been studied, see also~\cite{Endlich:2017tqa,Sennett:2019bpc,Bernard:2025dyh} for post-Newtonian methods and~\cite{Held:2021pht,Cayuso:2020lca,Held:2023aap,Cayuso:2023aht,Figueras:2024bba,Held:2025ckb} for non-linear time evolution. 

In the context of perturbative quantum gravity, non-local operators are associated with the finite part of the quantum effective action.
They arise from both semi-classical and quantum-gravity corrections, irrespective of whether one considers an effective or fundamental underlying theory.
At leading order in curvature, these corrections are captured by form factors $RF_R(\Box)R$, $S^{\mu\nu}F_S(\Box)S_{\mu\nu}$, and $C^{\mu\nu\rho\sigma}F_C(\Box)C_{\mu\nu\rho\sigma}$, with $\Box$ the covariant wave-operator. 
Here, we follow standard notation to decompose the curvature tensor into Ricci scalar~$R$, traceless Ricci tensor~$S_{\mu\nu}$, and Weyl tensor~$C_{\mu\nu\rho\sigma}$. 
Massive fields decouple at low energies~\cite{Appelquist:1974tg,Gorbar:2002pw,Gorbar:2003yt,Gorbar:2003yp} and one is left with a leading logarithmic correction, i.e., $F_i(\Box) = \log(\ell^2\Box)$, originating from the massless degrees of freedom. This structure was first derived in off-shell perturbation theory~\cite{Barvinsky:1985an,Barvinsky:1987uw,Barvinsky:1990up}
and later recovered in the context of effective field theory~\cite{Donoghue:1994dn,Bjerrum-Bohr:2002fji}.
Form factors such as $1/\Box$ or $1/(\Box + m^2)$ have also been considered as the starting point for phenomenological studies, for instance in the context of cosmology~\cite{Barvinsky:2003kg,Deser:2007jk,Ferreira:2013tqn,Woodard:2014iga,Maggiore:2014sia}, as they can reproduce the renormalization-group running of Newton's constant~\cite{Gorbar:2002pw}. The latter form factor may also be generated in the induced gravitational action by spontaneous symmetry breaking in curved spacetime~\cite{Gorbar:2003yp}.
\\

In general, specific approaches to quantum gravity give rise to more specific low-energy effective field theories characterized by distinct types of form factors. 
In asymptotically safe quantum gravity, first results for these form factors $F_i(\Box)$ have been obtained in~\cite{Christiansen:2014raa,Knorr:2018kog,Bosma:2019aiu,Knorr:2019atm,Knorr:2021niv,Bonanno:2021squ,Fehre:2021eob,Pawlowski:2025etp}, see also~\cite{Pawlowski:2023dda,DelPorro:2025wts} for first steps on charting the landscape of gravitationally localised objects.
In string theory, such form factors have been considered in~\cite{Fradkin:1985ys,Zwiebach:1992ie} and have been argued to take exponential forms, see, e.g.,~\cite{Calcagni:2013eua,Calcagni:2014vxa}. 
Proposals for a UV completion of gravity within the perturbative approach involve higher-derivative and non-local form factors already in the classical action, e.g.,~\cite{Modesto:2015ozb,Krasnikov:1987yj,Kuzmin:1989sp,Tomboulis:1997gg,Modesto:2011kw,Biswas:2011ar}.

In approaches to quantum gravity that feature a fundamental discreteness scale, the connection between ultraviolet physics and low-energy effective field theory is less obvious and, in fact, represents a key challenge.
Examples are causal set theory~\cite{Surya:2019ndm,Henson:2006kf} and loop quantum gravity~\cite{Ashtekar:2021kfp,Rovelli:1997yv,Sahlmann:2010zf} or spin-foam models~\cite{Perez:2012wv,Perez:2004hj,Baez:1999sr}. Candidate effective field theories for the continuum limit of spin foams have been derived taking into account the enlarged configuration space of the gravitational spin-foam path integral parametrized by area metrics~\cite{Dittrich:2021kzs,Dittrich:2022yoo,Dittrich:2023ava,Borissova:2022clg}. Area metrics~\cite{Schuller:2005yt,Schuller:2005ru,Punzi:2006nx,Punzi:2006hy,Schuller:2009hn,Ho:2015cza,Borissova:2024cpx,Bhattacharya:2025hag} feature massive non-metric degrees of freedom, in addition to the standard massless length-metric degrees of freedom~\cite{Borissova:2022clg,Borissova:2023yxs,Borissova:2025frj}.
While the original area-metric actions are local theories, effective length-metric actions can be obtained by integrating out these additional degrees of freedom. 
The result, starting from a local Lagrangian of second order in area-metric fluctuations and derivatives, are linearisations of non-linear quasi-local Einstein-Weyl actions of the type considered here~\cite{Borissova:2022clg,Borissova:2023yxs,Borissova:2025frj}.
The respective mass scale is set by the scale of the non-metric degrees of freedom. 
\\

Quasi-local form factors will be the starting point of our analysis. For concreteness, we will adopt the following terminology. We reserve the notion of non-local form factors to functions $F_i$ that are not analytic at $\Box =0$, and refer to functions $F_i$ that can be expanded as a series at $\Box = 0$ as quasi-local form factors. 
Concretely, we consider an exemplary quasi-local action that is quadratic in the Weyl curvature and analyse its effect on static spherically symmetric solutions. In particular, we determine local Frobenius solutions to this quasi-local action and compare them to previous results for the respective local action. 

The remainder of this article is structured as follows. In Section~\ref{Sec:ActionEOM} we introduce quasi-local Einstein-Weyl gravity and derive the covariant equations of motion after localisation of the action.  Subsequently we focus on their reduction in static spherical symmetry. In Section~\ref{Sec:WeakField} we analyse solutions in the weak-field regime. Section~\ref{Sec:Frobenius} is devoted to a detailed classification of Frobenius-series solutions around generic expansion points. We conclude with a discussion in Section~\ref{Sec:Discussion}.

\section{Action and equations of motion}\label{Sec:ActionEOM}

We  consider the following higher-derivative quadratic-curvature quasi-local Einstein-Weyl action,~\footnote{Here $m_{\text{Pl}}^2 = \frac{1}{8 \pi G} $ is the reduced squared Planck mass in Planck units $c=\hbar = 1$. In the following we set $m_{\text{Pl}}^2\equiv 1$.}
\be\label{eq:Action}
S\qty[g] =\frac{m_{\text{Pl}}^2}{2} \int\dd[4]{x} \sqrt{-g}\qty[R +\mu C_{\mu\nu\rho\sigma}\qty(\eta \Box - m^2)^{-1}C^{\mu\nu\rho\sigma}]\,,
\ee
where $\Box = g_{\mu\nu} \nabla^\mu \nabla^\nu$ is the covariant d'Alembert operator in signature convention $(-,+,+,+)$ and $\mu$ is a dimensionless coupling. The dimensionless parameter $\eta $ quantifies the amount of non-locality, whereas $m^2 > 0$ with dimension of mass squared introduces an additional mass scale originating from, e.g., a quantum-gravitational, high-energy completion.

The overparametrization in the second term of~\eqref{eq:Action} serves to distinguish different limits of this action. For $\mu \to 0$ the second term is absent and the action reduces to the classical Einstein-Hilbert action of general relativity. For $\eta \to 0$ we obtain the local Einstein-Weyl gravity as a special case of quadratic gravity~\cite{Stelle:1977ry,Salvio:2018crh}. For $m^2 \to 0$ the action becomes a genuinely non-local Einstein-Weyl action as a subclass of generalized non-local quadratic gravity~\cite{Conroy:2014eja}. Keeping both $\eta$ and $m^2$ generic for finite $\mu$ allows us to analyse the effects stemming from the covariant d'Alembert operator in combination with a mass term in the form of a quasi-local inverse operator between two Weyl curvatures.

\subsection{Localisation}
\label{sec:localisation}

To derive the equations of motion from~\eqref{eq:Action} we first localise the action by introducing an auxiliary tensor field $\psi_{\mu\nu\rho\sigma}$ with the same symmetries as the Weyl tensor and defined symbolically by 
\be\label{eq:Psi}
\psi_{\mu\nu\rho\sigma} \equiv - \qty(\eta \Box - m^2)^{-1}C_{\mu\nu\rho\sigma}\,.
\ee
This definition can be implemented at the level of the action by introducing a Lagrange multiplier $\lambda_{\mu\nu\rho\sigma}$ and considering the action
\be\label{eq:ActionLocalized}
S\qty[g,\psi,\lambda]  =\frac{m_{\text{Pl}}^2}{2} \int\dd[4]{x} \sqrt{-g}\qty[R -\mu C^{\mu\nu\rho\sigma}\psi_{\mu\nu\rho\sigma} + \lambda^{\mu\nu\rho\sigma}\qty(\qty(\eta \Box - m^2)\psi_{\mu\nu\rho\sigma} + C_{\mu\nu\rho\sigma})]\,,
\ee
such that the variation of this action with respect to the Lagrange multiplier gives back the local version of the defining relation~\eqref{eq:Psi},
\be\label{eq:PsiLocal}
\qty(\eta \Box - m^2)\psi_{\mu\nu\rho\sigma} = - C_{\mu\nu\rho\sigma}\,.
\ee
The equivalence of this equation to~\eqref{eq:Psi} requires specifying the homogeneous solution and choice of Green's function in the most general solution~\cite{Maggiore:2013mea,Foffa:2013vma,Maggiore:2014sia}. Here we will use the retarded Green's function and set the homogeneous solution to zero. In other words, we require that the operator acting on the left hand side in~\eqref{eq:PsiLocal} has a trivial kernel. This will allow us later to satisfy the boundary condition for $\psi$ that ensures a fall-off of this field at infinity for asymptotically flat spacetimes.
\\

The Lagrange multiplier can be integrated out from~\eqref{eq:ActionLocalized} to arrive at a reduced action which depends only on the metric and the field $\psi$,
\be\label{eq:ActionLocalizedReduced}
S\qty[g,\psi]  \equiv\frac{m_{\text{Pl}}^2}{2} \int\dd[4]{x} \sqrt{-g}\qty[R - \mu\qty( 2C^{\mu\nu\rho\sigma}\psi_{\mu\nu\rho\sigma} + \psi^{\mu\nu\rho\sigma}\qty(\eta\Box - m^2)\psi_{\mu\nu\rho\sigma})]\,.
\ee
In the following, we take this localised action as the starting point for our analysis of static and spherically-symmetric solutions.

\subsection{Equations of motion}

Varying the localised action~\eqref{eq:ActionLocalizedReduced} with respect to the field $\psi$, we obtain its equations of motion 
\be\label{eq:EOMPsi}
\mathcal{E}_{\mu\nu\rho\sigma}(g,\psi) \equiv  -\mu \qty[\qty(\eta \Box - m^2)\psi_{\mu\nu\rho\sigma} +  C_{\mu\nu\rho\sigma}] = 0\,.
\ee
Thus onshell for $\psi$ the action~\eqref{eq:ActionLocalizedReduced} reduces to~\eqref{eq:Action}. Varying~\eqref{eq:ActionLocalizedReduced} with respect to the metric and substituting the Ricci-Weyl decomposition for the Riemann tensor, we arrive at the equations of motion for the metric,
\be\label{eq:EOM}
\mathcal{E}_{\mu\nu}(g,\psi) \equiv  G_{\mu\nu} -  \mathcal{T}_{\mu\nu}(g,\psi) = 0\,.
\ee
Here $G_{\mu\nu} = R_{\mu\nu}-1/2 g_{\mu\nu}R$ is the Einstein tensor and the effective energy-momentum tensor is given by
\ba\label{eq:Teff}
\mathcal{T}_{\mu\nu} = -\mu\qty(\mathcal{T}_{\mu\nu}^{(0)} + \eta \mathcal{T}_{\mu\nu}^{(\eta)})\,,
\ea
where
\ba\label{eq:Teff0}
\mathcal{T}_{\mu\nu}^{(0)} &\equiv& -2 \nabla_{(\alpha}\nabla_{\beta)} \psi\indices{_\mu^\alpha_\nu^\beta}  - \frac{1}{2}m^2 g_{\mu\nu}\psi^{\alpha\beta\gamma\delta} \psi_{\alpha\beta\gamma\delta} + 4 m^2 \psi_{\mu}{}^{\alpha\beta\gamma}\psi_{\nu\alpha\beta\gamma}\nn\\
&{}& +g_{\mu\nu}C^{\alpha\beta\gamma\delta}\psi_{\alpha\beta\gamma\delta} - 6 C_{(\mu|}{}^{\alpha\beta\gamma}\psi_{|\nu) \alpha\beta\gamma} - 2 R^{\alpha\beta}\psi_{\mu\alpha\nu\beta} 
\ea
and
\ba\label{eq:TeffEta}
\mathcal{T}_{\mu\nu}^{(\eta)} &\equiv& -4\psi^{\alpha\beta\gamma\delta} \nabla_{\beta }\nabla_{(\mu}\psi_{\nu)\alpha\gamma\delta}  +4\psi_{(\mu|\alpha\beta\gamma} \nabla_{\delta }\nabla_{|\nu)}\psi^{\alpha\delta\beta\gamma}   \nn\\
&{}&- 4 \nabla_{(\mu |}\psi_{|\nu)\alpha\beta\gamma} \nabla_\delta \psi^{\alpha\delta\beta\gamma} +4 \nabla_{\delta} \psi_{(\mu|\alpha\beta\gamma} \nabla_{|\nu)}  \psi^{\alpha\delta\beta\gamma} \nn\\
&{}& - 4 \psi_{(\mu|\alpha\beta\gamma} \Box \psi_{|\nu)}{}^{\alpha\beta\gamma} + \nabla_\mu \psi_{\alpha\beta\gamma\delta}\nabla_\nu \psi^{\alpha\beta\gamma\delta} - \frac{1}{2} g_{\mu\nu} \nabla_\rho \psi_{\alpha\beta\gamma\delta} \nabla^{\rho}\psi^{\alpha\beta\gamma\delta}\,.
\ea
Equations~\eqref{eq:EOMPsi} and~\eqref{eq:EOM} define the full set of covariant equations of motion for the action~\eqref{eq:ActionLocalizedReduced}.

\subsection{Static spherically symmetric ansatz}

In the following we consider the equations of motion~\eqref{eq:EOMPsi} and~\eqref{eq:EOM} restricted to static and spherical symmetry. 
Depending on the choice of coordinates, the analysis of weak-field and generic Frobenius-series solutions can be considerably simplified.
Therefore, we shall use two different coordinate systems and metric ans\"atze.
\\

First we consider the metric line element parametrized in Schwarzschild spherical coordinates $(t,r,\theta,\phi)$  by two free functions $f(r)$ and $h(r)$, in the form
\be\label{eq:Metric}
\dd{s}^2 = -f(r)\dd{t}^2 + h(r)\dd{r}^2 + r^2 \dd{\Omega}^2\,,
\ee
where $\dd{\Omega}^2 =  \dd{\theta}^2+ \sin^2(\theta) \dd{\phi}^2$ is the area element on the unit two-sphere. To determine the form of the field $\psi_{\mu\nu\rho\sigma}$ in static spherical symmetry, we remind that this field has the same symmetries as the Weyl tensor. In particular it satisfies
\be\label{eq:PsiSymmetries}
\psi_{\mu\nu\rho\sigma} = - \psi_{\nu\mu\rho\sigma} = \psi_{\rho\sigma\mu\nu} \quad \,\,\,\text{and} \quad \,\,\, \psi_{\mu\alpha\beta\gamma} + \psi_{\mu\beta\gamma\alpha} +\psi_{\mu\gamma\alpha\beta} =0\,,
\ee
and is in addition traceless upon contracting any two of its four indices, as is the Weyl tensor. For a generic metric the latter can be written in Petrov notation as a $6\times 6$ symmetric matrix $C_{AB}$ where  $A,B,\dots = [\mu\nu] \in \{01,02,03,12,13,23\} $ label antisymmetric index pairs. 
Evaluated for the line element~\eqref{eq:Metric}, this matrix is diagonal and given by
\ba\label{eq:WeylSpherical}
C_{AB} = \mathcal{F } \cdot \mathcal{M}_{AB} \,,
\ea
where
\be\label{eq:FFunction}
\mathcal{F}\qty(f,f',f'',h,h';r) = \frac{1}{12}\qty[2 f'' - \frac{{f'}^2}{f} - \frac{f' h'}{h} - \frac{2f'}{r} +  \frac{2g h'}{r} +  \frac{4f}{r^2} -  \frac{4f h}{r^2}]\,,
\ee
and
\ba\label{eq:MAB}
\mathcal{M}_{AB} & = & \text{diag}\qty{1\,,\, -r^2 \frac{1}{2 h}\,,\, -r^2\sin^2(\theta) \frac{1}{2 h}\,,\,  r^2 \sin^2(\theta) \frac{1}{2 f}  \,,\, - r^4 \sin^2(\theta) \frac{1}{fh}}\,.
\ea
A prime denotes the derivative with respect to the coordinate $r$.
The non-zero components of the Weyl tensor are thus fully encoded in the diagonal entries of $\mathcal{M}_{AB}$ and related by the tracelessness condition $g^{\mu\rho}C_{\mu\nu\rho\sigma}=0$.
In view of the defining equation~\eqref{eq:EOMPsi} for $\psi_{\mu\nu\rho\sigma}$ and the structure of the Weyl tensor~\eqref{eq:WeylSpherical}, we can parametrize the field $\psi_{\mu\nu\rho\sigma}$ in the analogous form
\ba\label{eq:PsiAnsatz}
\psi_{AB}&=& \psi \cdot \mathcal{M}_{AB} \,,
\ea
where $\psi(r)$ is a scalar function depending only on the radial coordinate. It is straightforward to verify that the field $\psi_{\mu\nu\rho\sigma}$ defined in~\eqref{eq:PsiAnsatz} is traceless in each index pair and satisfies the cyclicity condition in~\eqref{eq:PsiSymmetries}. 

With the ans\"atze for the metric $g_{\mu\nu}$ in~\eqref{eq:Metric} and the field $\psi_{\mu\nu\rho\sigma}$ in~\eqref{eq:PsiAnsatz}, only the diagonal components of the field-equation tensors $\mathcal{E}_{\mu\nu}$ in~\eqref{eq:EOM} and $\mathcal{E}_{\mu\nu\rho\sigma}$ in~\eqref{eq:EOMPsi} (viewed as a $6\times 6$ matrix $\mathcal{E}_{AB}$) are non-vanishing.
Explicitly, these matrices take the form
\ba\label{eq:EOMg}
\mathcal{E}_{\mu\nu} & = & \begin{pmatrix}
    \mathcal{E}_{tt} & 0 & 0& 0 \\
    0 & \mathcal{E}_{rr}& 0 & 0 \\ 
    0 &0& \mathcal{E}_{\theta\theta} & 0 \\ 
    0 & 0&0 & \sin^2(\theta)\mathcal{E}_{\theta\theta}\\ 
\end{pmatrix} \,,
\ea
and
\ba\label{eq:EOMpsi}
\mathcal{E}_{AB} = \mathcal{E}_{trtr} \cdot \mathcal{M}_{AB}\,,
\ea
where $\mathcal{M}_{AB}$ is given in~\eqref{eq:MAB}. In particular, as a consequence of static spherical symmetry which implies that $\psi_{\mu\nu\rho\sigma}$ has to be of the form~\eqref{eq:PsiAnsatz}, there is only one independent equation of motion associated with $\mathcal{E}_{\mu\nu\rho\sigma}$, which we will take to be the $\mathcal{E}_{trtr}$ component.

Generically, the components of the field-equation matrices depend on the functions $f$, $h$ and $\psi$, their first and second derivatives, as well as on the coordinate $r$ explicitly. On-shell, i.e., on a solution to the equation of motion for the field $\psi$, the tensor $\mathcal{E}_{\mu\nu}$ is covariantly conserved,
\be\label{eq:BianchiIdentity}
\nabla^\mu \mathcal{E}_{\mu\nu} = 0\,.
\ee
 This statement is a consequence of the generalized Bianchi identities implied by general covariance of the action. For a tensor $\mathcal{E}_{\mu\nu}$ of the form~\eqref{eq:EOMg}, the only a priori non-vanishing component of the vector on the left-hand side in~\eqref{eq:BianchiIdentity} is the $r$ component,
\be\label{eq:BianchiIdentityrComponent}
\nabla^\mu \mathcal{E}_{\mu r} = \frac{f'}{2 f^2}\mathcal{E}_{tt}+ \frac{f'}{2 f h}\mathcal{E}_{rr}   + \frac{2}{h r} \mathcal{E}_{rr}
+ \qty(\frac{\mathcal{E}_{rr}}{h})' 
- \frac{2}{r^3}\mathcal{E}_{\theta\theta}\,.
\ee
The term $\mathcal{E}_{rr}'$ in this expression introduces derivatives of $f$, $h$ and $\psi$ up to third order. We can eliminate $\psi'''$ by taking the radial derivative of the equation of motion $\mathcal{E}_{trtr} = 0$ and substitute the result back into~\eqref{eq:BianchiIdentityrComponent}. The resulting expression involves derivatives of $f$, $h$ and $\psi$ only up to second order. Using therein  the equation of motion $\mathcal{E}_{trtr}= 0$ itself, to replace $\psi''$, leads to $\nabla^\mu \mathcal{E}_{\mu r} = 0$. In summary, we have verified explicitly that the conservation equation~\eqref{eq:BianchiIdentity} holds after using both the equation of motion for the field $\psi_{\mu\nu\rho\sigma}$ and its derivative. \\

Equation~\eqref{eq:BianchiIdentityrComponent} implies that there are only two independent equations of motion associated with $\mathcal{E}_{\mu\nu} = 0$. In the following we take these to be $\mathcal{E}_{tt} = 0$ and $\mathcal{E}_{rr} = 0$.
Altogether, we obtain a set of three algebraically independent coupled second-order non-linear differential equations for the functions $f$, $h$ and $\psi$, which we denote by
\ba\label{eq:EOMSpherical}
\mathcal{E}_{\text{Schwarzschild}} \equiv \qty{\mathcal{E}_{f}, \mathcal{E}_{h},  \mathcal{E}_{\psi} }\,.
\ea
Explicitly these equations of motion are given in Appendix~\ref{App:EOM}. They can be split into two parts: One part, denoted by a superscript $(\text{GR})$, remains after taking the limit $\mu \to 0$, when the original action reduces to the Einstein-Hilbert action of general relativity. 
The other part is proportional to the parameter $\mu$, and decomposes further into a contribution proportional to the non-locality parameter $\eta$ and denoted by a superscript $(\eta)$, as well as a contribution which remains in the limit $\eta \to 0 $ and is denoted by a superscript $(0)$. With this notation the equations of motion~\eqref{eq:EOMSpherical} can be expressed as
\ba
\mathcal{E}_f &\equiv& \mathcal{E}_f^{\text{GR}} - \mu \qty(\mathcal{E}_f^{(0)} + \eta  \mathcal{E}_f^{(\eta)}) = 0\,,\label{eq:Ef}\\
\mathcal{E}_h &\equiv & \mathcal{E}_h^{\text{GR}} - \mu \qty(\mathcal{E}_h^{(0)} + \eta  \mathcal{E}_h^{(\eta)}) = 0\,,\label{eq:Eh}\\
\mathcal{E}_{\psi} &\equiv & -\mu \qty(\mathcal{E}_\psi^{(0)} + \eta  \mathcal{E}_\psi^{(\eta)}) = 0\label{eq:EPsi}\,.
\ea
The equation of motion $\mathcal{E}_\psi$ has no GR contribution consistently with the absence of the field $\psi_{\mu\nu\rho\sigma}$ in general relativity. 

The equations $\qty{\mathcal{E}_{f}, \mathcal{E}_{h},  \mathcal{E}_{\psi} }\equiv 0$ are equivalent to the equations of motion obtained by first inserting the static spherically symmetric ansatz~\eqref{eq:Metric} and~\eqref{eq:PsiAnsatz} into the action~\eqref{eq:ActionLocalizedReduced} and subsequently taking the variation with respect to $f$, $h$ and $\psi$. According to the principle of symmetric criticality~\cite{Palais:1979rca,Deser:2003up,Frausto:2024egp}, this indicates consistency of our truncation ansatz to the invariant sector under the action of the compact rotation group.
\\

In addition to the above ansatz in Schwarzschild coordinates, 
a particularly convenient metric ansatz for the analysis of Frobenius solutions in Einstein-Weyl gravity is the conformal-to-Kundt one with coordinates $(u,\rho,\theta,\phi)$~\cite{Podolsky:2018pfe,Svarc:2018coe,Podolsky:2019gro}.
These are related to Schwarzschild coordinates $(t,r,\theta,\phi)$ by
\be\label{eq:CoordinateTransformation}
r = \Omega(\rho)\,\,\,\quad \text{and} \quad \,\,\, t = u - \int \frac{\dd{\rho}}{ \mathcal{H}(\rho)}\,.
\ee
The metric~\eqref{eq:Metric} in these coordinates takes the form
\be\label{eq:MetricKc}
\dd{s^2} = \Omega^2(\rho)\qty[\dd{\theta^2} + \sin^2\theta \dd{\phi^2} - 2 \dd{u}\dd{\rho} + \mathcal{H}(\rho)\dd{u^2}]\,,
\ee
and admits a two-parameter residual gauge freedom comprised of a shift of $\rho$ and a simultaneous rescaling of $\rho$ and $u$, namely,
\be
\label{eq:ResiGaugeCtK}
\rho \to \lambda \rho + \nu \, \qquad u \to \lambda^{-1} u \, .
\ee
The two metric functions $\Omega(\rho)$ and $H(\rho)$ are related to $f(r)$ and $h(r)$ in~\eqref{eq:Metric} via
\be\label{eq:FunctionTransformation}
f = - \Omega^2  \mathcal{H} \,\,\,\quad \text{and} \quad \,\,\, h = - \qty(\frac{\Omega}{\Omega'})^2  \frac{1}{\mathcal{H}}\,.
\ee
Although for the quasi-local version of the model the simplification of the field equations in Kundt coordinates is not as outstanding as for the local model, the metric~\eqref{eq:MetricKc} still has its merits. In fact, since the dependence on $\rho$ is only through the functions $\Omega$ and $\mathcal{H}$, the field equations form an autonomous system. This expedite calculations and facilitate the proof of existence of recursive relations for the series coefficients. Moreover, solutions that in Schwarzschild coordinates are non-Frobenius and seemingly unrelated, here appear together as subcases of a single family of Frobenius-series solutions in $\rho$.

Instead of explicitly carrying out the coordinate transformations~\eqref{eq:CoordinateTransformation} and~\eqref{eq:FunctionTransformation} in the equations of motion in Schwarzschild coordinates~\eqref{eq:EOMSpherical}, we will proceed by constructing a new ansatz for the field $\psi_{\mu\nu\rho\sigma}$ in Kundt coordinates and subsequently insert this ansatz together with the metric, expressed as in~\eqref{eq:MetricKc}, into the covariant equations of motion~\eqref{eq:EOMPsi} and~\eqref{eq:EOM}. To that end we observe that the, in this case non-diagonal, Petrov matrix for the Weyl tensor has only one algebraically independent component. Without loss of generality we consider this component to be
\be
C_{u\rho u\rho} = - \frac{1}{6}\Omega^2(\rho)\qty(2 + H''(\rho))\,,
\ee
where a prime denotes the derivative with respect to $\rho$. Consequently, we parametrize the field $\psi_{\mu\nu\rho\sigma}$ analogously as before in terms of one free scalar function ${\Psi}(\rho)$ by setting $ \psi_{u\rho u\rho} \equiv \Psi(\rho)$. Taking into account the generalized Bianchi identity, we are left with three independent equations of motion, which we take to be $\mathcal{E}_{uu} =0$ and $\mathcal{E}_{\rho\rho}=0$, as well as $\mathcal{E}_{u\rho u\rho} = 0$. These will be denoted by 
\be\label{eq:EOMSphericalKundtConformal}
\mathcal{E}_{\text{Kundt}} \equiv \qty{\mathcal{E}_{\Omega},\mathcal{E}_{\mathcal{H}},\mathcal{E}_\Psi}
\ee 
and can be expressed in the form
\ba
\mathcal{E}_\Omega &\equiv& \mathcal{E}_\Omega^{\text{GR}} - \mu \qty(\mathcal{E}_\Omega^{(0)} + \eta  \mathcal{E}_\Omega^{(\eta)}) = 0\,,\label{eq:EOmega}\\
\mathcal{E}_\mathcal{H} &\equiv & \mathcal{E}_\mathcal{H}^{\text{GR}} - \mu \qty(\mathcal{E}_\mathcal{H}^{(0)} + \eta  \mathcal{E}_\mathcal{H}^{(\eta)}) = 0\,,\label{eq:EH}\\
\mathcal{E}_{\Psi} &\equiv & -\mu \qty(\mathcal{E}_\Psi^{(0)} + \eta  \mathcal{E}_\Psi^{(\eta)}) = 0\label{eq:EPsiKundt}\,,
\ea
as given in Appendix~\ref{App:EOMKundtConformal}. These are coupled second-order differential equations for the functions $\Omega$, $\mathcal{H}$ and $\Psi$ which form an autonomous system as anticipated, i.e., they do not depend on the coordinate $\rho$ explicitly. 

\section{Weak-field regime}\label{Sec:WeakField}

In this subsection we will analyse solutions to the equations of motion in Schwarzschild coordinates~\eqref{eq:EOMSpherical} to linear order in an expansion around flat space. Such an approximation is suitable to describe the limit of a weak gravitational field at asymptotically large radii. Expanding the metric functions $f$ and $h$ in~\eqref{eq:Metric} and field $\psi$ in~\eqref{eq:PsiAnsatz} around their flat-space configuration, we write
\ba\label{eq:WeakFieldAnsatz}
f(r) &=& 1 + \delta a(r)\,,
\nonumber\\
h(r) &=& 1 + \delta b(r) \,,\\
\psi(r) &=& \delta c(r)\,.
\nonumber
\ea
The field $\psi$ has to vanish at leading order when the spacetime is described by the Minkowski metric with vanishing Weyl tensor. This is consistent with the assumption that solutions to the homogeneous equation in~\eqref{eq:PsiLocal} vanish.

Inserting the ansatz~\eqref{eq:WeakFieldAnsatz} into the field equations~\eqref{eq:EOMSpherical}  and expanding the result to first order in the perturbation parameter $\delta$ leads to
\ba\label{eq:EOMWeakFieldExpansion}
0&=& \qty[ b + r b' - \mu \qty(12 c + 20 r c' + 4 r^2 c'')]\delta+ \mathcal{O}\qty(\delta^2) \,,\nonumber\\
0  &=& \qty[b - r a' + \mu \qty(12 c + 4 r c')] \delta + \mathcal{O}\qty(\delta^2) \,,\\
0 &=& -\mu \qty[4 b + 12 m^2 r^2 c  + 2r a' - 2 r b' - 2 r^2 a'' + \eta\qty(72 c - 24 r c' -12 r^2 c'') ]\delta + \mathcal{O}\qty(\delta^2) \,.
\nonumber
\ea
We  now eliminate $b$ from the second equation and insert the result into the first and third equations. The thereby obtained second-order linear coupled differential equations for the functions $a$ and $c$ can be solved exactly in terms of four integration constants $C$, $C_{2,0}$ and $C_{2\pm}$. The most general solution to the linearized equations for the two metric functions $f$ and $h$ in the weak-field limit is given by
\ba
f(r) &=& 1 +C + \frac{C_{2,0}}{r} + C_{2+}\frac{e^{\tilde{m}_2r}}{r} +C_{2-} \frac{e^{-\tilde{m}_2r}}{r}\,,\label{eq:SolutionLinearizedf}\\
h(r) &=& 1 -  \frac{C_{2,0}}{r} - C_{2+} \frac{e^{\tilde{m}_2 r}}{2r}\qty(1 - \tilde{m}_2r) -  C_{2-} \frac{e^{-\tilde{m}_2r}}{2r}\qty(1 + \tilde{m}_2r)\label{eq:SolutionLinearizedh}\,.
\ea
The solution for the field $\psi$ is
\be\label{eq:SolutionLinearizedPsi}
\psi(r) = \frac{1}{m^2} \frac{C_{2,0}}{ r^3} + C_{2+} \frac{ e^{\tilde{m}_2r}}{8 \mu \tilde{m}_2^2 r^3}\qty(  3- 3 \tilde{m}_2r +\tilde{m}_2^2  r^2)  + C_{2-} \frac{ e^{-\tilde{m}_2 r}}{8 \mu   \tilde{m}_2^2 r^3}\qty( 3+3 \tilde{m}_2 r +  \tilde{m}_2^2 r^2)\,.
\ee
In the previous expressions we have introduced the parameter
\be\label{eq:m2Tilde}
\tilde{m}_2\equiv \sqrt{\frac{m^2}{2\mu + \eta}}\,.
\ee

\begin{figure}[t]
	\centering
	\includegraphics[width=0.8\textwidth]{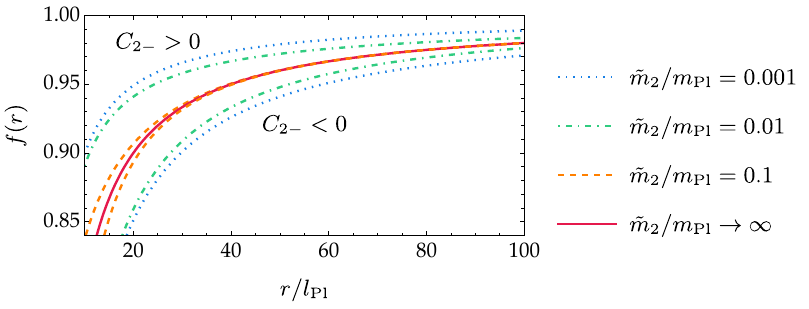}
	\caption{\label{Fig:WeakFieldLimitA} Metric function $f(r)$ in~\eqref{eq:SolutionLinearizedfAsymptFlat} for asymptotically flat solutions to the equations of motion in the weak-field regime, for different values of the effective mass parameter $\tilde{m}_2$ defined in~\eqref{eq:m2Tilde}. The integration constants are fixed to $M/m_\text{Pl}=1$ and $C_{2-}/m_{\text{Pl}}= \pm 1$. For $\tilde{m}^2 \to \infty$ the metric function asymptotes to its Schwarzschild counterpart represented by the red line.}
\end{figure}

The weak-field solution for the two metric functions~\eqref{eq:SolutionLinearizedf} and~\eqref{eq:SolutionLinearizedh} is written in a form that resembles expressions of weak-field solutions in the literature on quadratic gravity and more specifically local Einstein-Weyl gravity, obtained in the limit $\eta \to 0$ in the original action~\eqref{eq:Action}~\cite{Stelle:1977ry,Lu:2015psa,Perkins:2016imn,Bonanno:2019rsq,Silveravalle:2022wij,Bonanno:2022ibv}. In this limit $\tilde{m}_2^2 \to \frac{m^2}{2\mu} $ represents the squared mass of the spin-$2$ ghost appearing in the spectrum~\cite{Stelle:1977ry}. If $\mu<0$, this mass becomes imaginary and produces a non-asymptotically flat and spatially oscillating profile of the metric functions at large $r$, unless the corresponding coefficients $C_{2 \pm}$ are set to zero. A similar oscillatory behaviour would occur in the weak-field limit of the quasi-local system for generic values of the parameters $\mu$, $m^2$ and $\eta$, provided $2 \mu +\eta < 0$, in which case the radicand in~\eqref{eq:m2Tilde} is negative. In the following we will assume that this radicand is positive so that $\tilde{m}_2$ is real and positive. For the solutions to fall off and guarantee an asymptotically flat metric, we must set $C_{2+}= 0$. In addition we may fix the time parametrization at infinity by imposing $A(r)\to 1$ for $r\to \infty$ and thus setting $C=1$. Thereby the number of free parameters is reduced to two. These are $C_{2,0} =- 2M$, where $M$ can be identified with the Arnowitt-Deser-Misner mass of the Schwarzschild spacetime, and $C_{2-}$, which plays the role of a charge mediating a Yukawa interaction. This interaction can be attractive or repulsive depending on the sign of $C_{2-}$. The general form of asymptotically flat solutions to the weak-field equations is therefore
\ba
f(r) &=& 1  - \frac{2M}{r} +C_{2-} \frac{e^{-\tilde{m}_2r}}{r}\,,\label{eq:SolutionLinearizedfAsymptFlat}\\
h(r) &=& 1 +  \frac{2M}{r}  -  C_{2-} \frac{e^{-\tilde{m}_2r}}{2r}\qty(1 + \tilde{m}_2r)\,,\label{eq:SolutionLinearizedhAsymptFlat}
\ea
together with
\be\label{eq:SolutionLinearizedPsiAsymptFlat}
\psi(r) = -\frac{1}{m^2} \frac{2M}{ r^3} + C_{2-} \frac{ e^{-\tilde{m}_2 r}}{8 \mu   \tilde{m}_2^2 r^3}\qty( 3+3 \tilde{m}_2 r +  \tilde{m}_2^2 r^2)\,.
\ee
Figure~\ref{Fig:WeakFieldLimitA} shows the time-time component of the metric given by the function $f(r)$ in~\eqref{eq:SolutionLinearizedf} for fixed $M$ and $C_{2-}$ and different values of the parameter $\tilde{m}_2>0$. In the limit $\tilde{m}_2 \to \infty$, this function asymptotes to the lapse function of the Schwarzschild spacetime,
\ba
\lim_{\tilde{m}_2 \to \infty} f(r) = 1 - \frac{2M}{r}\,.
\ea
The latter is an exact solution to general relativity and local Einstein-Weyl gravity, but it is not a solution to the field equations of the quasi-local Einstein-Weyl theory, as we discuss in the next section.

In view of the definition~\eqref{eq:m2Tilde}, the limit $\tilde{m}_2 \to \infty$ can be achieved on the one hand by taking $m^2 \to \infty$. In the context of effective actions for the continuum limit of spin foams~\cite{Borissova:2022clg}, and more generally in area-metric gravity~\cite{Borissova:2023yxs,Borissova:2025frj}, this corresponds to a limit in which the non-metric degrees of freedom of the area metric become infinitely heavy and thereby decouple. In this limit the quadratic term in the Weyl tensor in the effective action for the length metric~\eqref{eq:Action} is effectively suppressed.
Notably, the limit $\tilde{m}_2 \to \infty$ can formally also be achieved taking $\mu \to -\frac{1}{2}\eta$. This special choice of couplings in the action~\eqref{eq:Action} leads to a ghost-free propagator for the spin-2 mode in an expansion around flat Minkowski background, which in particular does not exhibit additional poles beyond the massless graviton pole~\cite{Borissova:2022clg,Borissova:2023yxs,Borissova:2025frj}. Thus, one may view the absence of corrections to the weak-field regime of general relativity as a manifestation of the absence of additional degrees of freedom in this regime. Nevertheless, as the limit $\mu \to -\frac{1}{2}\eta$ is not well-defined in~\eqref{eq:m2Tilde}, the equations of motion for the non-linear theory described by this particular choice of parameters should be analysed separately.

\section{Frobenius-series solutions}\label{Sec:Frobenius}

In this section we will derive Frobenius-series solutions to the equations of motion with the conformal-to-Kundt metric ansatz~\eqref{eq:EOMSphericalKundtConformal}. An analogous analysis has been performed extensively  in the context of quadratic gravity, which contains local Einstein-Weyl gravity as a subclass~\cite{Stelle:1977ry,Holdom:2002xy,Lu:2015psa,Perkins:2016imn,Pravda:2016fue,Podolsky:2018pfe,Svarc:2018coe,Podolsky:2019gro}.

We expand the metric functions and the additional field as series in powers of $\rho$ around any fixed finite value $\rho_0$,
\ba
\label{eq:expCTK}
\Omega(\rho) &=& \Delta^n \sum_{i=0}^{\infty} A_{n+i} \Delta^i\,,\nonumber\\
\mathcal{H}(\rho) &=& \Delta^p \sum_{i=0}^{\infty} B_{p+i} \Delta^i\,,\\
\Psi(\rho) &=& \Delta^q \sum_{i=0}^{\infty} C_{q+i} \Delta^i\,,\nonumber
\ea
where
\ba
\Delta \equiv \rho - \rho_0\,.
\ea
The first coefficients $A_n$, $B_p$ and $C_q$ in the expansions are assumed to be non-zero, so that the leading behaviour of the solution is governed by the exponents $n$, $p$ and $q$ which are yet to be determined. The parameter $\rho_0$ can be any real constant, and its particular value has no physical meaning, as it can be redefined through a shift of the coordinate $\rho$. Solutions of this type will be denoted by 
\ba
[n,p,q]\,.
\ea

Analogously we will analyse solutions to the field equations in an asymptotic expansion around $\rho \to \infty$. To that end we assume that the metric functions and the additional field are expanded in negative powers of $\rho$, in the form
\ba
\label{eq:expCTK-asy}
\Omega(\rho) &=& \rho^n \sum_{i=0}^{\infty} A_{n-i} \rho^{-i}\,,\nonumber\\
\mathcal{H}(\rho) &=& \rho^p \sum_{i=0}^{\infty} B_{p-i} \rho^{-i}\,,\\
\Psi(\rho) &=& \rho^q \sum_{i=0}^{\infty} C_{q-i} \rho^{-i}\,.\nonumber
\ea
Solutions of this type will be denoted by
\ba
[n,p,q]^\infty\,.
\ea

In both cases, we restrict the considerations to expansions~\eqref{eq:expCTK} and~\eqref{eq:expCTK-asy} with integer steps $i=0,1,2,\ldots$ and with dominant exponents $n$, $p$ and $q$ that do not depend on the parameters of the model. It is possible that there exist more general solutions beyond the scope of this study. First, it may be that the theory admits additional solutions that are not in the form of Frobenius expansions in the investigated coordinate systems, i.e., neither in Schwarzschild nor Kundt coordinates. Second, it remains possible that there are special solutions for which the exponents
$n$, $p$ and $q$ are functions of the parameters of the model, namely, $\eta$ and $m$. 
Solutions of both types are known to exist in general models of quadratic and six-derivative gravity~\cite{Giacchini:2024exc,Giacchini:2025mlv}.

For completeness, we introduce a similar ansatz for the metric functions $f$ and $h$ in Schwarzschild coordinates, in terms of leading exponents $\alpha$ and $\beta$ which will allow us to physically interpret a given solution family found in Kundt coordinates. A particular class of solutions for the metric in conformal-to-Kundt form~\eqref{eq:MetricKc} thus translates into a corresponding class of solutions for the metric in Schwarzschild coordinates~\eqref{eq:Metric} as
\be
[n,p] \,\,\,\to \,\,\,(\alpha,\beta)_{r_0}\,,
\ee
with the label $r_0$ indicating the point around which the solution is expanded, i.e., $f(r)\sim (r-r_0)^\alpha$ and $h(r)\sim (r-r_0)^\beta$.

The physical interpretation of the point around which the solution is expanded also depends on the indicial structure. For example, from Eq.~\eqref{eq:CoordinateTransformation} it follows that $\Omega(\rho_0) = 0$ implies that $\rho_0$ corresponds to the origin $r=0$ in Schwarzschild coordinates. Thus, a solution with $n>0$ represents an expansion around the centre of the geometry. Analogously, if $n<0$, $\rho\to\rho_0$ corresponds to $r\to\infty$.  Similar reasoning applies to the expansions~\eqref{eq:expCTK-asy} around $\rho\to\infty$: in this case $n<0$ corresponds to expansions around $r=0$.
Specifically, the relation between the leading exponents of the expansions in the two coordinate systems is provided by
\be
\label{eq:alfabetaG}
    \alpha = 2 + \frac{p}{n}\, , \qquad \beta =  \frac{2-p}{n} \,,
\ee
for $n\neq 0$. 

In the particular case of $n=0$, both the expansions around $\rho=\rho_0$ and $\rho\to\infty$ correspond to expansions around a finite $r=r_0\neq 0$. The relation between $[n,p]$ and $(\alpha,\beta)_{r_0}$ in this case is more involved, as it depends on the order $N\in \lbrace 1,2,\ldots \rbrace$ of the first non-zero coefficient $A_N$ after $A_0$. In quadratic gravity, solutions with $A_1 = 0$ are typically associated with non-Frobenius solutions in Schwarzschild coordinates~\cite{Podolsky:2019gro}. Concerning the leading exponents, for the expansions~\eqref{eq:expCTK} one obtains
\be
\label{eq:alfabeta0}
    \alpha = \frac{p}{N} \,, \qquad \beta = - \frac{p + 2 (N-1)}{N}\,.
\ee
Since $\Omega(\rho_0)\neq 0$, $\mathcal{H}(\rho_0) = 0$ indicate a horizon, solutions $[0,p]$ with $p>0$ correspond to expansions around a horizon located at $\rho_0$  (or, $r_0\neq 0$ in Schwarzschild coordinates). On the other hand, solutions $[0,0]$ with $A_1 = 0$ are of type $(0,\beta)_{r_0}$ with $\beta<0$, thus $f(r_0)\neq0$ and $h(r_0)\to\infty$, which is a typical behaviour of a wormhole throat.

The cases described above are enough to cover the solutions we encounter in quasi-local Einstein-Weyl gravity.
Table~\ref{Tab:SolutionFamilies} summarizes our results and explicitly shows the existing solution families and their number of free parameters.
In Table~\ref{Tab:SolutionFamilies2} we compare the classes of solutions found here with those of  local Einstein-Weyl and generic quadratic gravity, as well as general relativity. 
Interestingly, we do not find any new solution classes $[p,q]$ in comparison to the ones that already occur in local Einstein-Weyl gravity. 
This can be explained by the observation that, for a conformal-to-Kundt metric, the terms in the field equations that are proportional to $\eta$ are of the same differential order of the others. Therefore, they cannot be the dominant ones at the leading order of the field equations. Instead, since we only consider solutions with coupling-independent exponents, they might act as additional constraints to the values of $p$ and $q$ that solve the indicial equations of Einstein-Weyl gravity. For this reason, we actually find less classes of solutions than the ones of the local model, as the solutions $[-1,3]^\infty$ and $[1,0]$ cannot satisfy the indicial equations related to the parameter $\eta$.

\begin{table}
\begin{tabular}{ |p{3.1cm}|p{4.3cm}|p{1.9cm}|p{6.3cm}| }
	\hline
	\centering $[n,p]$ or $[n,p]^\infty$ & \centering parameters & \centering free param. & \quad \quad \quad \quad \quad \,\,interpretation\\
	\hline
    $[-1,2]^\infty$  & $A_{-1}, B_1, B_0$ &  $3 \to 1$   & regular core  \\
	\hline
	$[0,0]_{A_1\neq 0}$  & $\qty{A_0, A_1, B_0, B_1, C_0, C_1}_1, \rho_0$ &  $6 \to 4$   & generic point  \\
	\hline
	$[0,0]_{A_1=0,A_2\neq 0}$  & $\qty{A_0, B_0, B_1, C_0, C_1}_1, \rho_0$ &  $5 \to 3$   & non-symmetric throat \\
	\hline
    $[0,0]_{A_1=0,A_2\neq 0,\,\text{even}}$  & $\qty{A_0, B_0,C_0}_1, \rho_0$ &  $3 \to 1$   & symmetric throat \\
	\hline
	$[0,0]_{A_{1,2}=0,A_3\neq 0}$  & $\qty{A_0, B_0, B_1, C_0, C_1}_2, \rho_0$ &  $4 \to 2$   & non-symmetric throat  \\
	\hline
    $[0,0]_{A_{1,2,3}=0,A_4\neq 0}$  &  $\qty{A_0, B_0, B_1, C_0, C_1}_3, \rho_0$ &  $3 \to 1$   & non-symmetric throat  \\
	\hline
	$[0,1]_{A_1\neq 0}$  & $A_0, B_1, C_0, \rho_0$ & $4 \to 2$   & Schwarzschild-like horizon  \\
	\hline
	$[0,1]_{A_1=0}$ & $\qty{A_0,C_0}_1, B_1, \rho_0$ &  $3 \to 1$   & horizon  \\
	\hline
	$[0,1]_{A_{1,2}=0}$ & $B_1,  \rho_0$ &  $2 \to 0$   & horizon \\
	\hline
	$[-1,2]$  & $A_{-1}, B_3, \rho_0$ & $3 \to 1$   & asymptotic corrections to Schwarzschild  \\
	\hline
\end{tabular}
\caption{Classes of solutions in quasi-local Einstein-Weyl gravity. Parentheses with an index $i$ in the list of parameters indicate the existence of $i$ constraints between them; for example, $\qty{A_0, B_0, B_1, C_0, C_1}_2$ indicates two constraints, so that among these parameters only three are independent. The counting of free parameters is shown in the third column, with an arrow indicating the reduction to the number of physical free parameters, after taking into account the residual gauge freedom of the metric.}\label{Tab:SolutionFamilies}
\end{table}

\begin{table}
\begin{tabular}{ |p{3.3cm}|p{4cm}|p{2.6cm}| p{0.7cm}| p{3cm}|p{0.7cm}|  }
	\hline
	\centering $[n,p]$ or $[n,p]^\infty$& \centering interpretation &\centering quasi-local EW &\centering EW &\centering quadratic gravity &  GR\\
	\hline
    \hline
	$[-1,3]^\infty$  &   Schwarzschild-like core &  \centering &  \checkmark & \checkmark &  \checkmark  \\
	\hline
	$[-1,2]^\infty$  &   regular core &  \centering \checkmark & \checkmark & \checkmark & \checkmark \\
	\hline
	$[1,0]$  &  2-2 hole core &  \centering &  \checkmark & \checkmark & \\
	\hline
    \hline
	$[0,1]_{A_1\neq 0}$   & Schwarzschild-like horizon &  \centering \checkmark & \checkmark & \checkmark & \checkmark \\
	\hline
	$[0,1]_{A_1=0}$  &  horizon &  \centering \checkmark & \checkmark & \checkmark &  \\
	\hline
	$[0,1]_{A_{1,2}=0}$  &  horizon &  \centering \checkmark &  & \checkmark & \\
	\hline
    \hline
	$[0,0]_{A_1\neq 0}$  &  generic point &  \centering \checkmark & \checkmark & \checkmark & \checkmark \\
	\hline
	$[0,0]_{A_1=0,A_2\neq 0}$   & non-symmetric throat &  \centering \checkmark & \checkmark & \checkmark &  \\
	\hline
	$[0,0]_{A_1=0,A_2\neq 0,\,\text{even}}$    & symmetric throat &  \centering \checkmark & \checkmark & \checkmark &  \\
	\hline
	$[0,0]_{A_{1,2}=0,A_3\neq 0}$   & non-symmetric throat &  \centering \checkmark & \checkmark & \checkmark &  \\
	\hline
    $[0,0]_{A_{1,2,3}=0,A_4\neq 0}$   & non-symmetric throat &  \centering \checkmark &  & \checkmark &  \\
	\hline
    $[0,0]_{A_{1,2,3}=0,A_4\neq 0,\,\text{even}}$   & symmetric throat &  \centering  &  & \checkmark &  \\
	\hline
    \hline
	$[-1,2]$   & asymptotic corrections to Schwarzschild &  \centering \checkmark & \checkmark & \checkmark & \checkmark \\
	\hline
\end{tabular}
\caption{Comparison of classes of Frobenius series solutions with coupling-independent exponents in quasi-local and local Einstein-Weyl gravity, quadratic gravity and general relativity. The solutions of the local higher-derivative models have been described for the first time in~\cite{Stelle:1977ry,Holdom:2002xy,Lu:2015psa,Podolsky:2019gro,Giacchini:2025mlv}.
}\label{Tab:SolutionFamilies2}
\end{table}

\subsection{Classes of solutions in an expansion in powers of $\Delta $}

As a first step we determine the possible indicial structure $[n,p,q]$ of the solutions by solving the indicial equations, i.e., the field equations at the lowest order. Assuming $\mu,\eta,m\neq 0$ and exponents $n$, $p$ and $q$ that are independent of those parameters, one may verify that the equation $\mathcal{E}_\mathcal{H} =0$ at zeroth order can only be solved if $n=0$ and $q \geqslant 0$, or if $n=-1$ and $q > -4$. The remaining equations then fix the admissible values of $p$, leading to the structures
\begin{equation}
    [0,1,q]\,, \qquad  [0,0,q]\, ,  \qquad [-1,2,q]\,.
\end{equation}
To determine possible values for $q$, one ought to check the expansions of the field equations at higher orders. In the following we shall discuss each of these families of solutions.

\subsubsection{Solutions $[0,1,q]$}
\label{Sec:01q}

The analysis of the field equations together with a solution ansatz in the form $[0,1,q]$ reveals that the only possibilities are $q\in\lbrace 0,1\rbrace$.
To solve the field equations order by order, one can solve  $\mathcal{E}_{uu} = 0$, $\mathcal{E}_{\theta\theta} = 0$ and $\mathcal{E}_\Psi = 0$. As a result, the other components will be automatically satisfied due to the Bianchi identities. 
At each order $N$, these three equations can be written in the form
\begin{align}
\label{eq:01sys}
\begin{multlined}
B_1 \left[A_0^8 (2 N+1) - 4 A_0^4 C_0 \mu  (2 N+1) +48 C_0^2 \eta  \mu  (N+1) \right] A_{N+1}  
- 2 A_0^5 C_0 \mu  (N+2) B_{N+2}
\\
+2 A_0 B_1 \mu  \left[A_0^4 (2 N+1)-6 C_0 \eta  (N+1)\right]  C_{N+1} 
+ \ldots
 =  0 \,,
\end{multlined}
\nonumber
\\
\begin{multlined}
4 B_1 (N+1) \left(A_0^8 +2 A_0^4 C_0 \mu +42 C_0^2 \eta  \mu \right)  A_{N+1}
+ A_0^9 (N+2) B_{N+2} 
\\
- 4 A_0 B_1 \mu  (N+1) \left(A_0^4+6 C_0 \eta \right) C_{N+1} 
+ \ldots 
=  0 \,,
\end{multlined}
\\
\mu \left[ 
24 B_1 C_0 \eta  (N+1)  A_{N+1} 
- A_0^5  (N+2)  B_{N+2}
-6 A_0 B_1 \eta   (N+1)  C_{N+1}
\right]  
+ \ldots 
=  0\,,
\nonumber
\end{align}
where the ellipses denote terms that depend on the coefficients $A_i, B_{i+1}, C_i$ with $i=0,\ldots,N$. 
This system has a unique solution for $A_{N+1}, B_{N+2}$ and $C_{N+1}$, provided that its determinant is non-zero, which yields the condition\footnote{There could exist solutions even if~\eqref{eq:det00} is not satisfied, but they would probably have a different structure of free parameters. We shall not consider this possibility here.}
\begin{equation}
\label{eq:det00}
    \mu A_0 B_1 \left[
    A_0^8 (\eta +2 \mu )+8 A_0^4 C_0 \eta  \mu +40 C_0^2 \eta  \mu ^2 \right] \neq 0 \,.
\end{equation} 
Since taking $C_0=0$ does not affect the existence of solutions to the system~\eqref{eq:01sys}, the family $[0,1,1]$ is contained as the particular cases of the above solution but with $C_0=0$. Therefore, we shall group both families $[0,1,0]$ and $[0,1,1]$ into a larger class labelled $[n,p]=[0,1]$ that is equivalent to $[0,1,0]$, but with the possibility of having $C_0=0$.

For $N=0$, the system~\eqref{eq:01sys} can be solved for the coefficients
\begin{equation}\label{eq:Sol01QLocal}
\begin{aligned}
  A_1 & =  -\tfrac{\eta  \left(A_0^4+4 C_0 \mu \right) \left(A_0^6+4 A_0^2 C_0 \mu +6 C_0^2 \mu  m^2\right)-2 A_0^8 C_0 \mu  m^2}{A_0 B_1 \left[A_0^8 (\eta +2 \mu )+8 A_0^4 C_0 \eta  \mu  +40 C_0^2 \eta  \mu ^2\right]}\,,
   \\
   B_2 & =  \tfrac{2 A_0^{14} (\eta -\mu )-2 A_0^{10} C_0 \mu  \left(3 A_0^2 m^2-2 \eta \right)+2 A_0^6 C_0^2 \eta  \mu  \left(3 A_0^2 m^2+18 \eta -20 \mu \right)-24 A_0^2 C_0^3 \eta  \mu ^2 \left(5 A_0^2 m^2-6 \eta \right)+216 C_0^4 \eta ^2 \mu ^2 m^2}{A_0^6 \left[A_0^8 (\eta +2 \mu )+8 A_0^4 C_0 \eta  \mu  +40 C_0^2 \eta  \mu ^2\right]},
   \\
   C_1 & =  -\tfrac{A_0^{14}+A_0^{10} C_0 \left(A_0^2 m^2+4 \eta +4 \mu \right)+2 A_0^6 C_0^2 \mu  \left(A_0^2 m^2+22 \eta \right)+8 A_0^2 C_0^3 \eta  \mu  \left(3 A_0^2 m^2+14 \mu \right)+168 C_0^4 \eta  \mu ^2 m^2}{A_0^2 B_1 \left[A_0^8 (\eta +2 \mu )+8 A_0^4 C_0 \eta  \mu  +40 C_0^2 \eta  \mu ^2\right]}\,,
\end{aligned}
\end{equation}
and subsequently for $A_2, B_3, C_2$ and so on.
Therefore, the solutions $[0,1]$ have four free parameters: $\rho_0$, $A_0$, $B_1$ and $C_0$. Taking into account the residual gauge freedom~\eqref{eq:ResiGaugeCtK}, the number of physical parameters characterizing such a solution is reduced to two.

Note that in the local limit $\eta\to 0$ the condition~\eqref{eq:det00} is trivial, as $\mu,A_0,B_1\neq 0$ by assumption. Moreover, the
above expressions reduce to
\begin{equation}
\label{eq:Sol01Local}
    A_1 = \frac{C_0 m^2}{A_0 B_1}\,, 
    \qquad B_2 = -\frac{3 C_0 m^2}{A_0^2}-1\,,
    \qquad C_1 = -\frac{A_0^2 C_0 \left(A_0^2 m^2+4 \mu \right)+A_0^6+2 C_0^2 \mu  m^2}{2 A_0^2 B_1 \mu }\,,
\end{equation}
and the solution for $A_1$ can always be inverted to trade the free parameter $C_0$ for $A_1$,
\begin{equation}
\label{eq:c072}
    C_0 = \frac{A_0 A_1 B_1}{m^2}\, .
\end{equation}
A similar inversion is more complicated if $\eta \neq 0$, owing to the non-linear relation between these parameters. This shows the existence of a constraint in the parameter space, if $A_1$ is taken as the free parameter instead of $C_0$.~\footnote{In general, compared to local Einstein-Weyl gravity, the quasi-local model seems to be more constrained, as already suggested by the restriction~\eqref{eq:det00}.}

The solutions in this family are regular at $\rho=\rho_0$, as one may verify by computing the Kretschmann scalar $R_{\mu\nu\alpha\beta}R^{\mu\nu\alpha\beta}$ to see that it is finite. 
In static spherical symmetry, this implies that all polynomial curvature scalars (without external covariant derivatives) are finite~\cite{Bronnikov:2012wsj}.
Moreover, they cannot be Ricci-scalar-flat. Even though it is possible to have
%
\be
    R = -\tfrac{24 C_0 \eta  \mu  \left[
    A_0^{10}
    +A_0^6 C_0 (A_0^2 m^2-3 \eta +4 \mu )
    +4 A_0^2 C_0^2 \mu  (4 A_0^2 m^2-3 \eta )
    -18 C_0^3 \eta  \mu  m^2\right]}{A_0^8 \left[A_0^8 (\eta +2 \mu )+8 A_0^4 C_0 \eta  \mu  +40 C_0^2 \eta  \mu ^2\right]} + O(\rho-\rho_0) \, 
\ee
vanishing as $\rho\to\rho_0$ --- for example, if $C_0=0$ ---
there is no choice of parameters that achieve $R=0$ identically, unless we consider  the local limit $\eta\to 0$.

From a physical point of view, the solutions $[0,1]$ correspond to expansions around horizons. According to Eq.~\eqref{eq:alfabeta0}, a general solution with $A_1\neq 0$ represents a Schwarzschild-like horizon, i.e., with a leading behaviour $(1,-1)_{r_0\neq 0}$ in Schwarzschild coordinates. On the other hand, the subfamily of solutions with $A_1=0$ have a distinctive interpretation as expansions around a non-Schwarzschild-like horizon, sometimes called an unusual horizon~\cite{Perkins:2016imn,Podolsky:2019gro}.
While in the local model this condition is achieved by simply fixing the free parameter $A_1=0$ (or, equivalently, $C_0=0$, see~\eqref{eq:Sol01Local}), in the quasi-local model this is not always possible and requires that a more complicated relation between the other parameters is satisfied. For example, fixing $C_0=0$ in~\eqref{eq:Sol01QLocal} yields
\begin{equation}
    A_1 = - \frac{A_0 \eta }{B_1 (2 \mu + \eta)}
    \neq 0
    \,.
\end{equation}
Specifically, from the solution for $A_1$ in~\eqref{eq:Sol01QLocal} it follows that the parameters of the solution must satisfy
\begin{equation}
\label{eq:constA1=0}
   \frac{\eta A_0 \left(A_0^4+4 C_0 \mu \right){}^2}{2 C_0 \mu  \left(A_0^8 -3 A_0^4 C_0 \eta -12 C_0^2 \eta  \mu \right)} = m^2\,,
\end{equation}
so that $A_1=0$. In particular, the requirement $m^2>0$ imposes 
constraints on the parameters $A_0$ and $C_0$ in relation to the parameters of the model. A solution with $A_1 = 0$ and $A_2\neq 0$ is related to the structure $(1/2,-3/2)_{r_0\neq 0}$ in Schwarzschild coordinates.

Interestingly enough, in the quasi-local model there is the possibility of having $A_2 = 0$, in addition to $A_1=0$, which relates to a solution with structure $(1/3,-5/3)_{r_0\neq 0}$ in Schwarzschild coordinates. This can be achieved if the parameters $A_0$ and $C_0$ satisfy
\begin{equation}
    \frac{A_0^8 \left(A_0^4-6 C_0 \mu \right)}{6 C_0 \left(A_0^8+6 A_0^4 C_0 \mu +8 C_0^2 \mu ^2\right)} = \eta \, , \qquad A_0^2 \left(\frac{A_0^4}{C_0^2 \mu }+\frac{96 \mu }{A_0^4+10 C_0 \mu }-\frac{12}{C_0}\right) = 6 m^2 \,.
\end{equation}
This cannot happen in the local model, as it would require $C_0 =  \tfrac{A_0^4}{6\mu} \neq 0$ and, at the same time, $C_0=0$ (see~\eqref{eq:c072} with $A_1=0$). Therefore, there is an extra type of unusual horizon allowed by the parameter $\eta$ that does not occur in the local Einstein-Weyl gravity.~\footnote{A general model of quadratic gravity which also contains the term $R^2$ in the action can also admit this type of solution~\cite{Giacchini:2025mlv}.} This solution has no free physical parameter, since both $A_0$ and $C_0$ are fixed by the constants of the model.

Finally, one may verify that the Schwarzschild spacetime is not present in this family of solutions if $\eta \neq 0$. Indeed, it would require $A_1=-A_0/B_1$, $A_2=A_0/B_1^2$, $B_2=2$ and $B_3=1/B_1$, which can only be satisfied if $\mu=0$ (general relativity) or 
$\eta = 0$ (Einstein-Weyl gravity). This conclusion also follows from the observation that there are no solutions in this class such that $R=0$ in the quasi-local model.

\subsubsection{Solutions $[0,0,q]$}

In the case of solutions $[0,0,q]$, the only possibilities are $q\in\lbrace 0,1,2\rbrace$. 
By using the generalized Bianchi identities, one can show that in order to solve the field equations order by order, it suffices to solve $\mathcal{E}_{uu} = 0$, $\mathcal{E}_{\theta\theta} = 0$ and $\mathcal{E}_\Psi = 0$ order by order, and $\mathcal{E}_{\rho\rho} = 0$ at the zeroth order.
To verify whether and how this can actually be done, let us first focus on the case $[0,0,0]$. 
At any given order $N$, the first triple of equations constitutes a linear system for the coefficients $A_{N+2}, B_{N+2}, C_{N+2}$,
\begin{align}
\label{00sys}
B_0 \left(A_0^8-4 \mu A_0^4 C_0+24 \mu \eta C_0^2 \right) A_{N+2} 
-\mu A_0^5 C_0 B_{N+2}
+2 \mu A_0 B_0 \left(A_0^4-3  \eta C_0 \right) C_{N+2} 
+ \ldots & = 0 \,,
\nonumber
\\
4 B_0 \left(A_0^8+2 \mu A_0^4 C_0+42 \mu \eta  C_0^2 \right)  A_{N+2}
+A_0^9 B_{N+2}
-4 \mu  A_0 B_0 \left(A_0^4+6 \eta C_0 \right)  C_{N+2} 
+ \ldots & = 0\, ,
\\ 
\mu \left[ 
24   \eta B_0 C_0 A_{N+2} 
-  A_0^5 B_{N+2}
- 6  \eta   A_0 B_0 C_{N+2}
\right]  
+ \ldots & = 0\,,
\nonumber
\end{align}
where the ellipses denotes terms that depend only on the coefficients $A_i, B_i, C_i$ with $i=0,\ldots,N+1$. This system has a unique solution provided that its determinant is non-zero, which precisely gives, again, the condition~\eqref{eq:det00}.
In addition, the constraint imposed by the equation $\mathcal{E}_{\rho\rho} = 0$ at the zeroth order reads
\begin{equation}
\label{eq:constr00}
\begin{split}
0 = & \, A_0^{10}+3 A_0^8 A_1^2 B_0
+ A_0^9 A_1 B_1
+2 \mu A_0^6 (B_1 C_1+2 C_0)
-4 \mu A_0^5 A_1  B_1 C_0 +6 \mu m^2 A_0^4 C_0^2 
\\
&
\, +6 \mu \eta A_0^2 B_0 C_1^2 
-48 \mu \eta A_0 A_1 B_0 C_0 C_1  
+ 132 \mu \eta A_1^2 B_0 C_0^2     \,.
\end{split}
\end{equation}

As setting $C_0=0$ and $C_1=0$ does not affect the above results, we can regard the solution classes $[0,0,1]$ and $[0,0,2]$ as subclasses of $[0,0,0]$ if, contrary to the original assumption, one permits the coefficients $C_0$ and $C_1$ to vanish.
Beyond this, it is also possible to set $C_2=0$ (by fixing $B_1=2A_0/A_1$), but this makes $\Psi(\rho) = 0$ --- which escapes the notation $[0,0,q]$. This last solution is known in closed form,
\begin{equation}
\label{Minkowski-CtK}
\Omega(\rho) = \frac{A_0^2}{A_0 - A_1 \Delta} = A_0 \sum_{i=0}^\infty \left( \frac{A_1 \Delta}{A_0} \right) ^i \,,
\qquad
\mathcal{H}(\rho) = -\frac{(A_0 - A_1 \Delta)^2}{A_1^2} \,,
\qquad
\Psi(\rho) = 0 \,,
\end{equation}
and corresponds to the Minkowski spacetime (see, e.g.,~\cite{Podolsky:2019gro}). Therefore, in a similar way as discussed for the class $[0,1]$, it is more convenient to refer to the larger class of solutions with indicial structure $[p,q]=[0,0]$ that contains the solutions $[0,0,0]$, $[0,0,1]$, $[0,0,2]$ and Minkowski.

At the lowest order $N=0$, the system~\eqref{00sys} can be solved for $A_2$, $B_2$ and $C_2$ (see Eq.~\eqref{eq:sol00N=0} in Appendix~\ref{App:formulae})
whereas Eq.~\eqref{eq:constr00} must be solved by a lower-order coefficient. 
Hence, a solution in the family $[0,0]$ is characterized by six free parameters: $\rho_0$ and five among $\lbrace A_0, A_1, B_0, B_1, C_0, C_1 \rbrace$, according to the constraint~\eqref{eq:constr00}. This is the same number of parameters that define the family $[0,0]$ in Einstein-Weyl gravity~\cite{Podolsky:2019gro}.

In principle, the free parameters $C_0$ and $C_1$ could be traded for $B_{2}$ and $B_3$, so that $\Psi$ would be fully determined once five coefficients among $A_{0,1}$ and $B_{0,1,2,3}$ are specified. Although this choice of basis of free parameters is natural in local Einstein-Weyl gravity, it is not convenient in the quasi-local model because of the constraints in the parameter space. The situation here is similar to the discussion on trading $C_0$ for $A_1$ in the solutions $[0,1]$ (see Sec.~\ref{Sec:01q}), but slightly more complicated due to the larger number of free parameters.

To elaborate on this, let us first recall the situation of the solutions $[0,0]$ in local Einstein-Weyl gravity. 
Taking the limit $\eta\to 0$ in Eq.~\eqref{eq:det00} yields simply $\mu A_0 B_0 \neq 0$, which in this context is trivially satisfied. 
Moreover, the structure of the last equation in~\eqref{00sys} simplifies considerably if $\eta = 0$ such that it can be solved for $C_N$ in terms of the parameters $A_{0,\ldots,N}$ and $B_{0,\ldots,N+2}$:
\begin{eqnarray}
\label{eq:00locC0}
C_0 & = & -\frac{A_0^2 (B_2 +1 )}{3 m^2}\,,
\\
\label{eq:00locC1}
C_1 & = &- \frac{2A_0 (B_2 + 1)}{3m^2} A_{1} - \frac{A_0^2}{ m^2} B_3\,,
\\
C_N & = & -\frac{2A_0 (B_2 + 1)}{3m^2} A_{N} - \frac{(N+2)(N+1) A_0^2}{6m^2} B_{N+2} + \ldots  \qquad (N \geqslant 2)\,,
\end{eqnarray}
where the ellipses denote terms depending on $A_{0,\ldots,N-1}$ and $B_{0,\ldots,N+1}$. Substituting these expressions into the expansions of the other components of the field equations, at zeroth order the three equations can be solved for $A_2$ and $B_4$ and one coefficient among $\lbrace A_0, A_1, B_0, B_1, B_2, B_3 \rbrace$, according to the constraint~\eqref{eq:constr00}, which now reads
\begin{equation}
\label{const1}
6 \mu  B_1 B_3 = 3 m^2 \left(A_0 A_1 B_1 + 3 A_1^2 B_0 + A_0^2\right)+2 \mu  \left(B_2^2-1\right)\,.
\end{equation}
For example, if $B_1 \neq 0$, it can be solved for $B_3$, whereas, if $B_1=0$, $B_3$ is a free parameter. At any higher order $N=1,2,\ldots$, the pair $\mathcal{E}_{uu} = 0$ and $\mathcal{E}_{\theta\theta} = 0$ can be solved for the coefficients $A_{N+2}$ and $B_{N+4}$. Therefore, we have a family with six free parameters: $\rho_0$ and five among $\lbrace A_0, A_1, B_0, B_1, B_2, B_3 \rbrace$. This is precisely the family $[0,0]$ described in~\cite{Podolsky:2019gro}. Notice that the subfamily defined by the condition $C_0 = 0$ corresponds to fixing $B_2=-1$ (see Eq.~\eqref{eq:00locC0}); subsequently, $C_1=0$ can be achieved by setting $B_3=0$ (see Eq.~\eqref{eq:00locC1}). 
Hence, the families $[0,0,q]$ with $q=0,1,2$ are contained in $[0,0]$ as $\Psi$ is fully determined once five coefficients out of $A_{0,1}$, $B_{0,1,2,3}$ are specified.

Returning now to the quasi-local model, as a simple example of change of basis of parameters, let us consider the subfamily of solutions with $C_0=0$ and assume that~\eqref{eq:constr00} is solved for $B_1$, i.e., the original parameters of the solution are $\lbrace \rho_0, A_0, A_1, B_0, C_0=0, C_1 \rbrace$. In this case, it can be shown that 
\begin{eqnarray}
\label{C00B1}
B_1 & = & -\frac{ A_0^8+3 A_1^2 A_0^6 B_0 +6  \mu  \eta B_0 C_1^2}{A_0^7 A_1 + 2 \mu  A_0^4 C_1}\,,
\\
\label{C00B2}
B_2 & = &  \frac{A_0^6 \eta  (2 A_0^2  + 3 A_1^2 B_0 ) - 2 \mu (A_0^8  -6 \eta A_0^3 A_1 B_0 C_1 - 3 \eta^2 B_0 C_1^2)}{A_0^8 (\eta+2 \mu)}\,.
\end{eqnarray}
To trade the parameters $C_0$ and $C_1$ for $B_1$ and $B_2$, one could invert the solution~\eqref{C00B1} for $B_1$,
to obtain
\begin{equation}
\label{C00C1}
C_1 = -\frac{ \mu A_0^4 B_1 \pm \sqrt{  \mu^2 A_0^8 B_1^2 - 6  \mu \eta  A_0^6 B_0 \left( A_0^2+ A_0 A_1 B_1 + 3 A_1^2 B_0\right)}}{6 \mu  \eta  B_0}\, ,
\end{equation}
if the radicand above is non-negative. (Notice that this restriction disappears in the local limit $\eta\to 0$ and the inversion is straightforward, as discussed above.) Under these circumstances, the basis $\lbrace \rho_0, A_0, A_1, B_0, C_0=0, C_1 \rbrace$ of free parameters changes to $\lbrace \rho_0, A_0, A_1, B_0, B_1 , B_2 \rbrace$ (with $B_2$ fixed by~\eqref{C00B2}--\eqref{C00C1}, which guarantees $C_0=0$). A similar discussion applies in the general case with $C_0\neq 0$, with more complicated conditions appearing to guarantee the positivity of radicands.

Regarding the physical interpretation, the class $[0,0]$ contains four distinct subfamilies. Three of them are similar to solutions in the local Einstein-Weyl gravity (whose interpretation was discussed in~\cite{Podolsky:2019gro}), while one does require $\eta \neq 0$. All of them have finite Ricci and Kretschmann scalars at $\rho=\rho_0$.
\begin{itemize}
    \item Family $[0,0]_{A_1\neq 0}$, corresponding to expansions around a generic regular point $\rho=\rho_0$. Such solutions have the maximum number of free parameters, that is, four physical parameters. In Schwarzschild coordinates, the leading terms have the structure $(0,0)_{r_0\neq0}$
    \item Family $[0,0]_{A_1=0,A_2\neq 0}$, corresponding to expansions around a wormhole-like throat and with three physical parameters.  If, in addition, $B_1=C_1=0$, the metric becomes even in $\Delta$ and the throat is symmetric. These families correspond to solutions in Schwarzschild coordinates with the structure $(0,-1)_{r_0\neq0}$. It is interesting to notice that, while in local Einstein-Weyl gravity the conditions $A_1=B_1=B_3=0$ are sufficient to guarantee a symmetric throat, in quasi-local Einstein-Weyl gravity it can happen that these conditions are satisfied with $C_1 \neq 0$, resulting in $A_3\neq 0$ and the throat being non-symmetric.
    \item Family $[0,0]_{A_{1,2}=0,A_3\neq 0}$, corresponding to expansions around a non-symmetric wormhole-like throat with two physical parameters. In Schwarzschild coordinates, the solution has the structure $(0,-4/3)_{r_0\neq 0}$.
    \item Family $[0,0]_{A_{1,2,3}=0,A_4\neq 0}$, corresponding to expansions around a non-symmetric wormhole-like throat with one physical parameter. To solve $A_2=A_3=0$ and the constraint~\eqref{eq:constr00} one has to fix three parameters, which is not possible if $\eta = 0$. Therefore, these types of solutions are not present in local Einstein-Weyl gravity~\cite{Podolsky:2019gro}, but exist in the more general  quadratic gravity~\cite{Giacchini:2025mlv}. In Schwarzschild coordinates they have the structure    $(0,-3/2)_{r_0\neq 0}$.
\end{itemize}

\subsubsection{Solutions $[-1,2,q]$}

There exists only one family of solutions of this type, with indicial structure $[-1,2,-1]$. Solutions can be obtained by solving, for example, the equations $\mathcal{E}_{uu} = 0$, $\mathcal{E}_{\rho\rho} = 0$ and $\mathcal{E}_\Psi = 0$ order by order, 
and as a consequence the other components will be automatically satisfied due to the Bianchi identities. At the first two orders, the field equations result in
\begin{equation}
    A_0=0 \,, \qquad B_2=-1\,, \qquad  C_{-1}= -\frac{A_{-1}^2 B_3}{m^2}\,,
\end{equation}
leaving $A_{-1}$ and $B_{3}$ as free parameters.
Afterwards, at each order we obtain a linear system $ M_N \textbf{x}_N  = \boldsymbol{\Phi}_N$, to be solved for the coefficients $A_{N-1}$, $B_{N+2}$, $C_{N-2}$, where $N=2,3,\ldots$, 
\ba\label{eq:Matrix-1-2-1}
	M_N = \begin{pmatrix}
		-\frac{2 B_2 (N-1) N}{A_{-1}} & N-1 & 0 \\
		\frac{4 B_2 N}{A_{-1}} & N-1 & 0  \\ 
		0 & \mu  (N+1) (N+2) & \frac{6 \mu  m^2}{A_{-1}^2}  
	\end{pmatrix} \, ,
    \qquad
    \textbf{x}_N =
    \begin{pmatrix}
		A_{N-1} \\
		B_{N+2}  \\ 
		C_{N-2}
	\end{pmatrix}\,,
\ea
and $\boldsymbol{\Phi}_N$ depends on $A_{-1,\ldots,N-2}$, $B_{2,\ldots,N+1}$ and $C_{-1,\ldots,N-3}$. This system has a unique solution provided that 
\begin{equation}
\label{eq:det-12}
    \det \qty(M_N) = \frac{12 \mu  m^2 N \left(N^2-1\right)}{A_{-1}^3} \neq 0 \,.
\end{equation}
Unlike the analogous condition~\eqref{eq:det00} for the solutions $[0,1]$ and $[0,0]$, Eq.~\eqref{eq:det-12} does not depend on $\eta$ and is always satisfied for both local and the quasi-local Einstein-Weyl gravity (but not in the fully non-local limit $m^2\to 0$). Therefore, the solutions in the family $[-1,2,-1]$ are characterized by three free parameters, $\rho_0$, $A_{-1}$ and $B_3$.
Its first few terms read
\begin{eqnarray}
        \Omega(\rho) & = & \frac{A_{-1}}{\Delta} +\frac{3 B_3^2 \eta  \mu }{7 A_{-1}^3 m^4} \Delta^5 - \frac{60 B_3^2 \eta  \mu  (\eta + 3 \mu  )}{A_{-1}^5 m^6} \Delta^7 + O\qty(\Delta^8) \,,
        \nonumber
        \\
        \mathcal{H}(\rho) &=& - 
        \Delta^2 + B_3 \Delta^3 - \frac{162 B_3^2 \eta  \mu }{7 A_{-1}^4 m^4} \Delta^8 + \frac{138  B_3^3 \eta  \mu }{7 A_{-1}^4 m^4} \Delta^9 + O\qty(\Delta^{10})\,,
        \\
        \Psi(\rho) & = &- \frac{A_{-1}^2 B_3}{m^2 \Delta} +  \frac{3 B_3^2 \eta  }{m^4} \Delta^2 + \frac{72 B_3^2 \eta (\eta + 3 \mu  )}{A_{-1}^2 m^6}  \Delta^4 + O\qty(\Delta^{5})\,.
        \nonumber
\end{eqnarray}

According to the discussion related to Eq.~\eqref{eq:alfabetaG},  since $p=-1$ is negative, this solution corresponds to asymptotic expansions in powers of $1/r$ in Schwarzschild coordinates, with indicial structure $(0,0)_\infty$, namely, 
\begin{eqnarray}
    f(r) & = & a_0 \left[ 1 +\frac{a_1}{r} 
    +\frac{24 a_1^2 \eta  \mu }{m^4 r^6}
    +\frac{21 a_1^3 \eta  \mu }{m^4 r^7}
    + \frac{360 a_1^2 \eta  \mu  (\eta +3 \mu )}{m^6 r^8}    
    \right] + O\qty(r^{-9}) \, ,
    \nonumber
    \\
    \frac{1}{h(r)} & = & 1 
    +\frac{a_1}{r}
    +\frac{18 a_1^2 \eta  \mu }{m^4 r^6}
    +\frac{15 a_1^3 \eta  \mu }{m^4 r^7}
    + \frac{1440 a_1^2 \eta  \mu  (\eta +3 \mu )}{m^6 r^8} + O\qty(r^{-9}) \, ,
    \\
    \psi(r) & = & 
    \frac{a_0 a_1}{m^2 r^3}
    +\frac{3 a_0 a_1^2 \eta }{m^4 r^6}
    +\frac{72 a_0 a_1^2 \eta  (\eta + 3 \mu  )}{m^6 r^8}
    + O\qty(r^{-9}) \, .
    \nonumber
\end{eqnarray}
The only physical parameter is $a_1$, which can be interpreted as mass, while $a_0$ can be absorbed by a rescaling of the time coordinate.
We note that in general the metric functions do not obey the relation $f(r) \propto h(r)^{-1}$.
In the local limit $\eta \to 0$ the solution reduces to Schwarzschild. Here, it can be interpreted as asymptotic corrections to the Schwarzschild geometry in the regime $r\to \infty$.

It is interesting to notice that leading-order corrections of order $r^{-6}$ to the Schwarzschild geometry at large distances have been found also in six-derivative gravity~\cite{Giacchini:2025gzw,Daas:2024pxs}. 
This order in asymptotic $r^{-n}$ corrections is in agreement with the lower bound of $n\geqslant 6$ on local (or quasi-local) corrections obtained in~\cite{Knorr:2022kqp}.

\subsection{Classes of solutions in an expansion in powers of $\rho^{-1}$}

Following the same procedure outlined in the previous section, we determined the possible indicial structures $[n,p,q]^\infty$ for asymptotic solutions of the form~\eqref{eq:expCTK-asy}, as $\rho \to \infty$. In this case, the indicial equations are the field equations at the highest order. The only solution found, assuming that the exponents $n$, $p$ and $q$ do not depend on the couplings of the model, is $[-1,2,-6]^\infty$.

The solutions in the class $[-1,2,-6]^\infty$ can be obtained by solving the equations $\mathcal{E}_{uu} = 0$, $\mathcal{E}_{\rho\rho} = 0$ and $\mathcal{E}_\Psi = 0$ order by order. The corresponding system is underdetermined at the first three orders of expansion in powers of $r^{-1}$, resulting in
\begin{equation}
\label{eq:asumpreto}
    A_{-2} = \frac{1}{2} A_{-1} B_1\,,
    \qquad
    A_{-3} = \frac{1}{3} A_{-1} \left(B_0+B_1^2\right) \,,
    \qquad B_2 = -1\,,
    \qquad C_{-6} = -\frac{A_{-1}^4 \left(B_1^2+4 B_0\right)}{240 \mu }\,,
\end{equation}
where the parameters $A_{-1}$, $B_1$ and $B_0$ remain free. At higher orders, it is always possible to solve the three equations for three coefficients. Indeed, the system can be cast in the form $ M_N \textbf{x}_N  = \boldsymbol{\Phi}_N$, where $N=3,4,\ldots$ is related to the order of the expansion, and
\ba\label{eq:Matrix-1-2-6inf}
	M_N = \begin{pmatrix}
		2 N A_{-1}^3 & -A_{-1}^4 & 4 \mu  (N+3) \\
		4 N A_{-1}^3 & -(N+1)A_{-1}^4  & - 4 \mu  (N+3)  \\ 
		0 & - \mu  (N-2) (N-1) A_{-1}^4  & 6 \eta  \mu  (N-2) (N+3)  
	\end{pmatrix} \,,
    \qquad
    \textbf{x}_N =
    \begin{pmatrix}
		A_{-(N+1)} \\
		B_{-N+2}  \\ 
		C_{-(N+4)}
	\end{pmatrix}\,.
\ea
Also, $\boldsymbol{\Phi}_N$ depends on $A_{-1,\ldots,-N}$, $B_{2,\ldots,-N+3}$ and $C_{-6,\ldots,-N-3}$.
Hence, it can be uniquely solved for the coefficients $A_{-(N+1)}$, $B_{-N+2}$, $C_{-(N+4)}$ if $\det \qty(M_N) \neq 0$, that is,
\begin{equation}
\label{eq:det-126}
    \mu A_{-1}  N (N-1) (N-2) (N+3) (\eta +2 \mu) \neq 0 \,,
\end{equation}
which is always satisfied for the models and solutions considered here.
Therefore, the solutions in this family have three free parameters, $A_{-1}$, $B_1$ and $B_0$, but only one is physical, due to the residual gauge freedom~\eqref{eq:ResiGaugeCtK}. 
Choosing the gauge $A_{-1}=-1$ and $B_1=0$, the first terms read
\begin{eqnarray}
\Omega(\rho) & = & -\frac{1}{ \rho} -\frac{B_0}{3  \rho^3} - \frac{B_0 \left[ 30 \mu m^2 + B_0 ( 5 \eta^2    + 756  \mu  \eta + 1440 \mu^2 ) \right]}{3600  \mu   ( \eta + 2 \mu ) \rho^5} + O\qty(\rho^{-6})\,,
\nonumber
\\
\mathcal{H}(\rho) &=& - \rho^2 + B_0 + \frac{B_0 \left[ 30  \mu  m^2  + B_0 \eta  ( 7 \eta + 40 \mu  ) \right]}{900 \mu  ( \eta + 2 \mu ) \rho^2} + O\qty(\rho^{-3})\,,
\\
\Psi(\rho) &=& -\frac{B_0}{60 \mu  \rho^6} - \frac{B_0 \left[ 15 m^2 + B_0 ( 413 \eta + 800 \mu   )\right]}{12600 \mu  ( \eta + 2 \mu ) \rho^8} + O\qty(\rho^{-9})\,.
\nonumber
\end{eqnarray}
It is straightforward to verify that in the local limit $\eta \to 0$ we recover the corresponding solution $[-1,2]^\infty$ of Einstein-Weyl gravity found in~\citep{Podolsky:2019gro}.

The Ricci and Kretschmann scalars
\begin{eqnarray}
    R & = & \frac{(B_1^2+4 B_0)^2 \eta }{240 A_{-1}^2 \mu \, \rho^2} + O\qty(\rho^{-3}) \, ,
    \nn\\
    R_{\mu\nu\alpha\beta}R^{\mu\nu\alpha\beta} & = &
    (B_1^2+4 B_0)^2 \left[ \frac{1}{6 A_{-1}^4} + \frac{ 60 A_{-1}^2 \mu  m^2+(B_1^2+4 B_0) (3 \eta ^2+39 \eta  \mu +40 \mu ^2)}{1080 A_{-1}^4 \mu  ( \eta + 2 \mu ) \, \rho^2} \right] + O\qty(\rho^{-3})
    \nonumber
\end{eqnarray}
are finite in the limit $\rho \to \infty$. Thus there is no curvature singularity. Moreover, these invariants vanish for $B_0=-B_{1}^2/4$ (or equivalently, $C_{-6}=0$, see Eq.~\eqref{eq:asumpreto}), in which case $\Psi = 0$ and the solution reduces to the Minkowski spacetime~\citep{Podolsky:2019gro}; this is the only possibility for the solution to be Ricci-scalar flat. 

Since $\Omega(\rho)\to 0$ for $\rho\to\infty$, from Eq.~\eqref{eq:CoordinateTransformation} it follows that the solutions in this class correspond to expansions around the origin ($r=0$) in Schwarzschild coordinates, with indicial structure $(0,0)_{0}$. We can express the expansions of the functions $f$, $h$ and $\psi$ for this regular solution family in Schwarzschild coordinates as follows,
\ba
f(r) &=& a_0 \qty[1 + b_2 r^2 +  \frac{\qty(240 \mu^2 + 76 \mu \eta - 9\eta^2)b_2 + 10 \mu m^2 }{200 \mu ( \eta + 2 \mu )}b_2 r^4] +\mathcal{O}\qty(r^6) \,, \label{eq:ABPsiSolutionApproximateA}\\
h(r) &=& 1 + b_2 r^2 +  \frac{\qty(120\mu^2 + 28 \mu \eta - 3\eta^2)b_2 + 10 \mu m^2 }{100 \mu ( \eta + 2 \mu )}b_2 r^4 +\mathcal{O}\qty(r^6)\,, \label{eq:ABPsiSolutionApproximateB}\\
\psi(r) &=&  \frac{1}{20 \mu}b_2  r^2 + \frac{\qty(320\mu + 147\eta )b_2 + 5 m^2}{1400\mu ( \eta + 2 \mu )} b_2 r^4 +\mathcal{O}\qty(r^6)\,. \label{eq:ABPsiSolutionApproximatePsi}
\ea
Let us note that in the limit $\eta \to 0$, associated with local Einstein-Weyl gravity, the above expressions for the metric functions $f$ and $h$ coincide with expressions found in~\cite{Lu:2015psa,Perkins:2016imn,Holdom:2016nek} once we make the appropriate translation of couplings at the level of the action~\eqref{eq:Action}. Here we find for quasi-local Einstein-Weyl gravity the same number two of free parameters (a global time rescaling parameter $a_0$, and $b_2$) as in local Einstein-Weyl gravity for regular static spherically symmetric solutions at the origin. Deviations between the local and quasi-local versions of Einstein-Weyl, indicated by the appearance of the parameter $\eta$, occur only at orders $r^4$ and above. Notably, this is the only Frobenius solution family at $r=0$ in Schwarzschild coordinates that we find in quasi-local Einstein-Weyl gravity. In particular, the singular Frobenius solutions families $(-1,1)_0$ and $(2,2)_0$ that exist in quadratic gravity are not present.

\section{Discussion}\label{Sec:Discussion}

We have constructed static spherically symmetric solutions to quasi-local Einstein-Weyl gravity as Frobenius-series expansions in Kundt coordinates and with coupling-independent exponents.
To do so, we have parametrised quasi-local Einstein-Weyl gravity by (i) a dimensionless beyond-GR coupling $\mu$, (ii) a non-locality parameter $\eta$ and (iii) mass scale $m$: For $\mu\rightarrow0$, the action reduces to general relativity; for $\eta\rightarrow0$, the theory reduces to local Einstein-Weyl gravity; and in the formal limit of $m\rightarrow0$ the action becomes genuinely non-local.
We have then localised the action by introducing an auxiliary field with the same symmetries as the Weyl tensor and derived the static spherically symmetric equations of motion in Kundt coordinates. Expanding in a Frobenius series about a generic point $\rho=\rho_0$, we have systematically constructed \emph{all} admissible families of solutions and shown how they relate to Schwarzschild coordinates. The result is summarised in~\cref{Tab:SolutionFamilies}. To contrast the effect of the non-locality parameter $\eta$, we have compared our results in quasi-local Einstein-Weyl gravity (henceforth, the quasi-local theory) to previous results in local Einstein-Weyl gravity and, more generally, quadratic gravity~\cite{Stelle:1977ry,Lu:2015cqa,Lu:2015psa,Podolsky:2018pfe,Svarc:2018coe,Podolsky:2019gro,Held:2022abx,Giacchini:2025mlv} (henceforth, the local theory).

We caution that we cannot exclude the existence of solutions, which, either do not admit for a Frobenius expansion in the investigated coordinates, or for which the Frobenius exponents are coupling dependent, see also~\cite{Giacchini:2024exc,Giacchini:2025mlv}. Further, the formal limit of $m\rightarrow0$, in which the theory becomes truly non-local, is non-trivial and requires separate investigation, as well as the particular cases in which the conditions~\eqref{eq:det00} and~\eqref{eq:det-126} are not satisfied. 
\\

With these cautionary remarks in mind, our main result is striking: While singular solutions exist in the local theory, we find that these seem to be absent in the quasi-local theory.
Herein, the term ``regular'' refers to the finiteness of curvature scalars --- in particular, the Ricci and Kretschmann scalars --- around the expansion point.
Put differently, the non-locality seems to regularise singularities.

More generally, we can make the following statements about the solution space:
\begin{itemize}
    \item The Schwarzschild spacetime is not a solution. 
    \item In the flat asymptotic limit, i.e., around $r\rightarrow\infty$, the non-local corrections to the Schwarzschild spacetime scale like $1/r^{6}$. We highlight that this is the same order in $\mathcal{O}(1/r)$ than those coming from the tower of local curvature corrections, i.e., those arising from cubic-curvature terms~\cite{deRham:2020ejn,Giacchini:2025gzw,Daas:2024pxs}. 
    \item All of the Frobenius classes of solutions found here deform continuously to the corresponding solutions of the local theory and are regular. While some of the regular solutions are absent in local Einstein-Weyl gravity, all of them have a counterpart in quadratic gravity. In particular, this includes expansions around horizons and the regular core at Schwarzschild coordinate~$r=0$.
    \item In contrast, while local theories exhibit solutions with singular cores --- both in local Einstein-Weyl gravity or in quadratic gravity --- the quasi-local theory does not have singular solutions.
\end{itemize}
On a technical level, the differences in presence/absence of solutions arise since
the quasi-local model has one parameter more than the local model. On the one hand, this extra parameter allows for more general sub-classes of solutions within the classes $[0,0]$ and $[0,1]$ (the same sub-classes exist for a generic quadratic gravity model, which also has an extra parameter). On the other hand, the extra parameter adds new constraints to the solutions around $r=0$, which eventually exclude singular solutions from the quasi-local model.

Concerning the solution classes that are shared by the local and the quasi-local models, it is interesting to notice that the solutions of the latter seem to be more constrained than the ones in the same class in the local model, although the solutions are characterized by exactly the same number of parameters. This can be interpreted as if non-locality imposed certain constraints between the free parameters of the solutions and the couplings of the model. As a result, depending on the values of the couplings, there might exist forbidden regions in the parameter space of the solution; this situation does not happen with the solutions in the same classes in the local model. Alternatively, this observation might suggest that our choice of free parameters is not optimal, and the constraints could be artifacts from the localization procedure, which involves introducing an extra field.  Ultimately, to understand the  physical consequence of those constraints, it would be necessary to identify the physical meaning of the parameters. This can be very complicated, however, and  might require knowing the global structure of the solution.

It is also worth noting that although the same classes of solutions have the same number of free parameters in both local and quasi-local versions of Einstein-Weyl gravity, the actual solutions and the interpretation of their parameters can be very different. For example, the local model admits a 2-parameter family of black hole solutions in the class $[0,1]$, with parameters that can be related to the position of the horizon and to the non-triviality of the Bach tensor~\cite{Podolsky:2018pfe,Podolsky:2019gro}. When the so-called Bach parameter vanishes, one is left with the Schwarzschild black hole. In contrast, Schwarzschild is not a solution in the quasi-local model and all the solutions in the family $[0,1]$ have a non-vanishing Bach tensor. Whether this reveals the existence of a constraint in the range of the Bach parameter, or calls for a new interpretation of the parameter, is left as an open question.
\\

While the absence of singular Frobenius solutions is promising, the question of whether quasi-local Einstein-Weyl gravity admits regular black holes remains open. To construct complete regular black-hole solutions, one would have to connect the Frobenius expansions of the regular core, the horizon, and the asymptotic region. In this regard, the family of solutions with a regular core has only one physical parameter, $b_2$ (see Eqs.~(\ref{eq:ABPsiSolutionApproximateA}, \ref{eq:ABPsiSolutionApproximateB})), which should be related to the position of the horizons and, at the same time, ensure asymptotic flatness. If the latter requires fine-tuning a parameter (as in the case of the Schwarzschild-Bach black holes of quadratic gravity~\cite{Lu:2015cqa}), then such regular black holes would be the exception rather than the rule. Moreover, it is curious that solutions of the type $[0,2,q]$, corresponding to expansions around a double horizon, were not found. Indeed, since a regular and asymptotically flat black hole must have an even number of horizons, one might expect --- although we see no strict necessity --- that the extremal black hole (i.e., when two horizons degenerate into a double horizon) would be a solution as well. In any case, our result suggests that, if the quasi-local theory admits regular black-hole solutions, their extremal limit is not captured by the present Frobenius expansion and is hence non-trivial.
It would thus be interesting to numerically construct global solutions as has been done in quadratic gravity~\cite{Lu:2015cqa,Kokkotas:2017zwt,Held:2022abx}.

The analytical results of the present work can be a starting point to the numerical search for global solutions, e.g., by defining boundary conditions for the shooting method. In this regard, one ought to expect more difficulties than in quadratic gravity, as here we have a system of three coupled differential equations. In addition, as explained before, the phase space of solutions seems to be more constrained than in the local Einstein-Weyl gravity. The numerical analysis is far from trivial, and even if no global regular black hole solution is found, it would be difficult to prove, by means of numerical calculations, that they do not exist. 
Nevertheless, one may anticipate obstructions to the construction of global asymptotically flat regular solutions based on the local Frobenius series solutions found here purely from a parameter counting. In fact, the regular solution family at the core features only one physical parameter, $b_2$, which encapsulates all the available freedom in a numerical shooting towards asymptotic infinity. Regardless of whether such a shooting would take into account intermediate local solutions, in any asymptotically flat spacetime with a finite ADM mass, one may expect a required nontrivial  fine-tuning of the dependence on any free Frobenius solution parameters on the mass, down to the regular core. It is difficult to imagine such a mass-dependence of the inverse-square-length regularisation  parameter $b^2$ in the absence of an additional dimensionful scale.
On the other hand, singular spacetimes may be found 
in a numerical search of solutions, 
requiring further analytical studies to understand their local structure, e.g., in the form of Puiseux series, and their number of free parameters. We stress, however, that our results already prove that there exists no solution with the same type of singularity of those of Einstein-Weyl gravity (i.e., with a Schwarzschild-Bach or the so-called 2-2 hole type of singular radial core).
\\

If global solutions can be constructed, further study should address their dynamical formation and stability. 
For such dynamical questions, it is important to treat the initial data for the fiducial field $\Psi$ consistently, i.e., such that it removes the homogeneous solution and adheres to the respective choice of Green's function used when localising the quasi-local action (see~\cref{sec:localisation}).
Following prior work in scalar-tensor theories, see, e.g.,~\cite{Kovacs:2020pns,Kovacs:2020ywu,Ripley:2019aqj,Ripley:2020vpk,East:2020hgw,Corman:2024vlk,Lara:2025kzj,AresteSalo:2025sxc,Corman:2025wun}, in quadratic gravity~\cite{Noakes:1983xd,Held:2021pht,Held:2023aap,East:2023nsk,Held:2025ckb,May:2025arz}, and in the local tower of gravitational EFT corrections more generally~\cite{Figueras:2024bba}, this would entail (i) formulating a well-posed time evolution, i.e., a locally well-posed initial value problem, (ii) studying linear and nonlinear stability of the respective static spherically-symmetric solutions, and (iii) investigating the nonlinear endpoint of potential instabilities.
In this context, it is interesting to note that the covariant equations of motion may be ``regularised'' (following the prescription in~\cite{Figueras:2024bba}) by adding suitable terms, i.e., $\alpha\left(R^{\mu\nu}\Box R_{\mu\nu} - \frac{1}{2}R\Box R\right)$, to the localised action. 
As discussed in~\cite{Figueras:2024bba}, these terms will introduce further fiducial fields.
In local theories, see~\cite{Deffayet:2025lnj,Figueras:2025wtx,Held:2025fii} for instructive examples in scalar field theories and~\cite{Held:2023aap} for the nonlinear decoupling of such massive modes in quadratic gravity, these additional fiducial fields can be effectively decoupled by choosing $\alpha$ such that their masses are sufficiently large.
We emphasize that these additional fiducial fields are distinct from $\Psi$.
The fiducial field $\Psi$ itself is evolved according to a standard hyperbolic wave equation, see~\cref{eq:EOMPsi}.
Hence, we expect that any initial data that is consistent with the localisation, i.e., removes the respective homogeneous solution, will continue to describe the respective solution during evolution.
More generally, the combination of ``localisation'', ``regularisation'', and suitable choice of initial data could thus present a prescription of how to formulate a well-posed time evolution for generic quasi-local/non-local theories such as the one considered here. We will make this explicit in future work.
\\

Finally, it is worth mentioning that if the quasi-local Einstein-Weyl gravity is considered in the perspective of EFT, then the correspondence of our solutions and an EFT solution is not automatic. In fact, within this interpretation, the mass scale of the ghost mode marks the cutoff --- or at least the onset of a breakdown --- of the derivative expansion, and a physical EFT interpretation of a given branch requires checking that the derivative/curvature expansion remains ordered when evaluated on the local solution. Under the assumption of a joint curvature-derivative (i.e., power-counting) expansion of the EFT, this can be assessed by determining the curvature scalars, schematically denoted as $\mathcal{R}_i$, determining their mass dimension $d_i$, and then checking whether $|\mathcal{R}_i| < m^{d_i}$. For example, for the expansions around the regular core (see Eqs.~\eqref{eq:ABPsiSolutionApproximateA},~\eqref{eq:ABPsiSolutionApproximateB}), this criterion can be evaluated explicitly: The free parameter $b_2$ has mass dimension two and sets the leading central curvature scale. Direct evaluation of the non-vanishing curvature scalars at the regular centre therefore gives $|b_2| < m^2$ (up to numerical factors of order unity).
Whether or not the regular core is parametrically within the regime of validity of the EFT expansion thus depends on the parameter $b_2$.
\\

Overall, our results provide insight into how non-localities may play an important role in a more thorough understanding of the effects of quantum corrections to the effective action
and on its classical solutions --- such as the avoidance of singularities compared to general relativity. The regularity of onshell configurations, and hence finiteness of the onshell action, may be connected with a finite contribution in saddle-point approximations to the full path integral for quantum gravity, in addition to the possibility that offshell singular configurations may be efficiently suppressed through destructive interference~\cite{Borissova:2020knn,Giacchini:2021pmr,Borissova:2023kzq,Borissova:2024hkc}.
In combination with future work on obtaining the low-energy quantum effective action and respective on-shell solutions as considered here, this provides a promising route towards a more complete understanding of singularity resolution from first principles.
\\

\begin{acknowledgments}
	
We thank Bianca Dittrich and Andrew Tolley for discussions. We also thank the anonymous referee for helpful comments regarding an EFT interpretation and the construction of global solutions.
J.B.~is supported by STFC Consolidated Grant ST/X000575/1 and Simons Investigator Award~690508.
B.L.G.~acknowledges financial support by the Primus grant PRIMUS/23/SCI/005 from Charles University and the support from the Charles University Research Center Grant No.~UNCE24/SCI/016.
	
\end{acknowledgments}

\appendix
\section{Equations of motion}\label{Sec:Appendix}

\subsection{Equations of motion in Schwarzschild coordinates}\label{App:EOM}

The equations of motion~\eqref{eq:EOMSpherical} can be split into a part corresponding to the general relativity (GR) limit $\mu \to 0$ of the original quasi-local Einstein-Weyl action~\eqref{eq:Action} and its reduced localized version~\eqref{eq:ActionLocalizedReduced} together with another part proportional to the parameter $\mu$ which stands in front of the Weyl-curvature squared modification in the original action~\eqref{eq:Action}. The second part decomposes further into a contribution proportional to the non-locality parameter $\eta$ and denoted by a superscript $(\eta)$ as well as a contribution which remains in the local Einstein-Weyl limit $\eta \to 0 $ of the original action~\eqref{eq:Action} and is denoted by a superscript $(0)$. With this notation the  equations of motion~\eqref{eq:EOMSpherical} are explicitly given by

\ba
\mathcal{E}_f &\equiv& \mathcal{E}_f^{\text{GR}} - \mu \qty(\mathcal{E}_f^{(0)} + \eta  \mathcal{E}_f^{(\eta)}) = 0\,,\\
\mathcal{E}_h &\equiv & \mathcal{E}_h^{\text{GR}} - \mu \qty(\mathcal{E}_h^{(0)} + \eta  \mathcal{E}_h^{(\eta)}) = 0\,,\\
\mathcal{E}_{\psi} &\equiv & - \mu \qty(\mathcal{E}_\psi^{(0)} + \eta  \mathcal{E}_\psi^{(\eta)}) = 0\,,
\ea
where
\ba
\mathcal{E}_{f}^{\text{GR}} &=& f^4 h^3 \qty(r h'+h^2-h)\,,\\
\mathcal{E}_f^{(0)} &=& fh \Big(r fh\qty(11 r \psi f' h'-2 h \qty(-3 m^2 r \psi^2+4 r f' \psi'+3 \psi \qty(r f''+3 f'))) +9 r^2 h^2 \psi f'^2 \nn\\
&{}&+2 f^2 \big(5 r^2 \psi h'^2-r h \qty(5 r h' \psi'+\psi \qty(2 r h''+13 h'))+2 h^2 \qty(r \qty(r \psi''+5 \psi')+3 \psi )\big)\Big)\,,\quad\quad \\
\mathcal{E}_f^{(\eta)} &=& -24 r^2 h^2 \psi^2 f'^2 - 6 f^2 \Big(6 r^2 \psi^2 h'^2+h^2 \qty(r^2 \psi'^2+2 r \psi \qty(r \psi''+2 \psi')-6 \psi^2)- \nn\\
&{}& r h \psi \qty(7 r h' \psi'+2 \psi \qty(r h''+2 h'))\Big) + 6 r fh \psi \big(h \qty(2 r \psi f''+5 r f' \psi'+4 \psi f') \nn\\
&{}&-6 r \psi f'h'\big)
\ea
and
\ba
\mathcal{E}_{h}^{\text{GR}} &=& f^4 (h-1) h^4-r f^3 h^4 f'\,,\\
\mathcal{E}_h^{(0)} &=& -f h^2 \Big(2r f h \qty(-3 m^2 r \psi^2+r \psi f''-r f' \psi'-5 \psi f')+r^2 \psi f' h'+r^2 h \psi f'^2 \nn\\
&{}&+2 f^2 \qty(2 h \qty(r \psi'+3 \psi )-r \psi h')\Big)\,,\\
\mathcal{E}_h^{(\eta)} &=&  -12 r^2 h^2 \psi^2 f'^2- 6f^2 \Big(4 r^2 \psi^2 h'^2-h^2 \qty(r^2 \psi'^2-2 r \psi \qty(r \psi''+2 \psi')+18 \psi^2) \nn\\
&{}&-r h \psi\qty(3 r h' \psi'+2 \psi \qty(r h''+2 h'))\Big)  + 6r f h  \psi \big(h \qty(2 r \psi f''+r f' \psi'+4 \psi f') \nn\\
&{}&-2 r \psi f' h'\big)
\ea
and
\ba
\mathcal{E}_\psi^{(0)} &=& fh^2 \big(r f \qty(r f' h'-2 h \qty(-6 m^2 r \psi+r f''-f'))+r^2 h f'^2 +f^2 \qty(-2 r h'+4 h^2-4 h)\big)\,,\quad \quad \\
\mathcal{E}_\psi^{(\eta)} &=& -18 r^2 h^2 \psi f'^2 - 6 f^2 \Big(5 r^2 \psi h'^2 - r h \qty(5 r h' \psi'+2 \psi \qty(r h''+2 h')) \nn\\
&{}& - 2 h^2 \qty(6 \psi-r \qty(r \psi''+2 \psi'))\Big)+ 6r f h \qty(h \qty(2 r \psi f''+3 r f' \psi'+4 \psi f')-4 r \psi f' h' )\,.
\ea
The equation of motion for $\mathcal{E}_\psi$ is has no GR contribution consistently with the absence of the field $\psi_{\mu\nu\rho\sigma}$ in general relativity.

\subsection{Equations of motion in Kundt coordinates}\label{App:EOMKundtConformal}

The equations of motion~\eqref{eq:EOMSphericalKundtConformal} can be written as
\ba
\mathcal{E}_\Omega &\equiv& \mathcal{E}_\Omega^{\text{GR}} - \mu \qty(\mathcal{E}_\Omega^{(0)} + \eta  \mathcal{E}_\Omega^{(\eta)}) = 0\,,\\
\mathcal{E}_\mathcal{H} &\equiv & \mathcal{E}_\mathcal{H}^{\text{GR}} - \mu \qty(\mathcal{E}_\mathcal{H}^{(0)} + \eta  \mathcal{E}_\mathcal{H}^{(\eta)}) = 0\,,\\
\mathcal{E}_{\Psi} &\equiv & -\mu \qty(\mathcal{E}_\Psi^{(0)} + \eta  \mathcal{E}_\Psi^{(\eta)}) = 0\,.
\ea
where
\ba
\mathcal{E}^{\text{GR}}_\Omega &=& \Omega^{10} - \mathcal{H} \Omega^8 \Omega'^2 + 
\Omega^9 \mathcal{H}' \Omega' + 2 \mathcal{H} \Omega''\,,\\
\mathcal{E}^{(0)}_\Omega &=&  4 \Omega^5 \big(\Psi \big(\mathcal{H}' \Omega' +2 \mathcal{H}\Omega''\big)+4 \mathcal{H}\Psi' \Omega'\big) +2\Omega^6 \big(\Psi \mathcal{H}''-\mathcal{H}' \Psi'-2 \mathcal{H} \Psi'' \big)\nn\\
&-&6 \Psi \Omega^4 \big(4 \mathcal{H} \Omega'^2 -m^2 \Psi\big)\,,
\\
\mathcal{E}^{(\eta)}_\Omega &=&  12 \Psi \Omega  \big(\Omega \big(\mathcal{H}' \Psi' + \mathcal{H} \Psi''\big)-10 \mathcal{H}\Psi'  \Omega'\big)-6\Psi^2 \big(8 \Omega \big(\mathcal{H}' \Omega'+\mathcal{H} \Omega''\big)-34 \mathcal{H}\Omega'^2\big)\nn\\
&+&6\mathcal{H}\Omega^2 \Psi'^2
\ea
and
\ba
\mathcal{E}^{\text{GR}}_\mathcal{H} &=& \Omega^9\Omega''-2 \Omega^8 \Omega'^2\,,\\
\mathcal{E}^{(0)}_\mathcal{H} &=& -2\Omega^6\Psi''+4 \Omega^5 \big(2 \Psi' \Omega'+\Psi \Omega''\big)-12 \Psi \Omega^4 \Omega'^2\,,
\\
\mathcal{E}^{(\eta)}_\mathcal{H} &=&  -48 \Psi \Omega \Psi'\Omega' +6\Omega^2 \Psi'^2+132 \Psi^2 \Omega'^2\,,
\ea
and
\ba
\mathcal{E}^{(0)}_\Psi &=& \Omega^6 \big(\mathcal{H}''+2\big)+6 m^2 \Psi \Omega^4\,,
\\
\mathcal{E}^{(\eta)}_\Psi &=& -12\Omega \big(2\Psi \big(\mathcal{H}' \Omega' +\mathcal{H} \Omega''\big)+3 \mathcal{H} \Psi' \Omega' \big)+6\Omega^2 \big(\mathcal{H}' \Psi' +\mathcal{H}\Psi'' \big)+36\mathcal{H}\Psi \Omega'^2 \,.
\ea
Note that $\mathcal{E}_\mathcal{H}$ is independent of $\mathcal{H}$.

\section{Coefficients $A_2$, $B_2$ and $C_2$ for the solutions $[0,0]$}
\label{App:formulae}

The solution of the system~\eqref{00sys} with $N=0$ for the coefficients $A_2$, $B_2$ and $C_2$ is:
\begin{small}
{\allowdisplaybreaks 
\begin{subequations}
\label{eq:sol00N=0}
\begin{eqnarray} 
A_2 & = & \{ 
2 A_0^8 \mu  [
A_0^6
-5 A_1 A_0^5 B_1
+3 A_0^4 (A_1^2 B_0+2 C_0 m^2)
+2 A_0^2 \mu  (B_1 C_1+2 C_0)
-4 A_1 A_0 B_1 C_0 \mu 
+6 C_0^2 \mu  m^2 ] 
\nonumber
\\
&& 
-3 A_0^4 \eta  [
A_0^{10}
+A_1 A_0^9 B_1
-A_1^2 A_0^8 B_0
-A_0^6 \mu(2 B_1 C_1  -8 C_0 )
+8 A_1 A_0^5 \mu  (2 B_1 C_0+B_0 C_1)
\nonumber
\\
&& 
-2 A_0^4 C_0 \mu  (2 A_1^2 B_0-3 C_0 m^2)
+4 A_0^2 \mu ^2 (4 C_0^2-2 B_1 C_1 C_0-3 B_0 C_1^2)
+32 A_1 A_0 C_0 \mu ^2 (3 B_1 C_0
\nonumber
\\
&& 
+7 B_0 C_1)
-24 C_0^2 \mu ^2 (25 A_1^2 B_0-C_0 m^2)
]
+
18 B_0 \eta ^2 \mu  (22 A_1^2 C_0^2-8 A_0 A_1 C_1 C_0+A_0^2 C_1^2) (4 C_0 \mu +A_0^4)
\}
\nonumber
\\
&& 
\times \{
12 A_0^5 B_0 [A_0^8 (\eta +2 \mu ) +8 A_0^4 C_0 \eta  \mu +40 C_0^2 \eta  \mu ^2]
\}^{-1} \, ,
\\
B_2 & = & \{
A_0^8 \eta  [
A_0^{10}
-A_1 A_0^9 B_1
+2 A_0^6 \mu  (C_0-B_1 C_1)
+6 A_1 A_0^5 \mu  (B_1 C_0+2 B_0 C_1) 
+6 A_1^2 A_0^4 B_0 C_0 \mu 
\nonumber
\\
&& 
-4 A_0^2 C_0 \mu ^2 (8 C_0-B_1 C_1)
-8 A_1 A_0 B_1 C_0^2 \mu ^2
-108 C_0^3 \mu ^2 m^2]
-2 A_0^{16} \mu  (A_0^2+3 C_0 m^2)
\nonumber
\\
&& 
+ 18 A_0^4 C_0 \eta ^2 \mu  [
A_0^6 C_0
-A_1 A_0^5 (B_1 C_0+4 B_0 C_1)
+13 A_1^2 A_0^4 B_0 C_0
-2 A_0^2 \mu  (B_1 C_1 C_0-B_0 C_1^2-2 C_0^2)
\nonumber
\\
&& 
+4 A_1 A_0 C_0 \mu  (B_1 C_0-6 B_0 C_1)
+2 C_0^2 \mu  (3 C_0 m^2+50 A_1^2 B_0)
]
-108 B_0 C_0^2 \eta ^3 \mu ^2 (22 A_1^2 C_0^2-8 A_0 A_1 C_1 C_0
\nonumber
\\
&& 
+A_0^2 C_1^2)
\}
\times \{
A_0^{10} [A_0^8 (\eta +2 \mu ) +8 A_0^4 C_0 \eta  \mu +40 C_0^2 \eta  \mu ^2]
\}^{-1} \, ,
\\
C_2 & = & \{
A_0^8 [
-2 A_0^{10}
+A_1 A_0^9 B_1
-3 A_0^8 C_0 m^2
-A_0^6 \mu \left(4 B_1 C_1  +6 C_0 \right)
-2 A_1 A_0^5 \mu  \left(B_1 C_0-12 B_0 C_1\right)
\nonumber
\\
&& 
-30 A_1^2 A_0^4 B_0 C_0 \mu +
4 A_0^2 C_0 \mu ^2 \left(B_1 C_1+2 C_0\right)
-8 A_1 A_0 B_1 C_0^2 \mu ^2
+12 C_0^3 \mu ^2 m^2
]
- 3 A_0^4 \eta  [
A_0^{10} (B_1 C_1
\nonumber
\\
&& 
+2 C_0)
-2 A_1 A_0^9 (B_1 C_0+3 B_0 C_1)
+4 A_1^2 A_0^8 B_0 C_0
+2 A_0^6 C_0 \mu  (2 B_1 C_1+11 C_0)
-2 A_1 A_0^5 C_0 \mu  (3 B_1 C_0
\nonumber
\\
&& 
+28 B_0 C_1)
+2 A_0^4 C_0^2 \mu  (6 C_0 m^2+59 A_1^2 B_0)
+4 A_0^2 C_0 \mu ^2 (3 B_1 C_1 C_0-3 B_0 C_1^2+14 C_0^2)
\nonumber
\\
&& 
+8 A_1 A_0 C_0^2 \mu ^2 (7 B_1 C_0+8 B_0 C_1)
-12 C_0^3 \mu ^2 (30 A_1^2 B_0-7 C_0 m^2)
]
+
36 B_0 C_0 \eta ^2 \mu  (22 A_1^2 C_0^2
\nonumber
\\
&& 
-8 A_0 A_1 C_1 C_0+A_0^2 C_1^2) (7 C_0 \mu +A_0^4)
\}
\times \{
6 A_0^6 B_0 [A_0^8 (\eta +2 \mu ) +8 A_0^4 C_0 \eta  \mu +40 C_0^2 \eta  \mu ^2]
\}^{-1} \, .
\end{eqnarray} 
\end{subequations}
}
\end{small}

\bibliographystyle{jhep}
\bibliography{references}

@article{Barvinsky:1985an,
    author = "Barvinsky, A. O. and Vilkovisky, G. A.",
    title = "{The Generalized Schwinger-Dewitt Technique in Gauge Theories and Quantum Gravity}",
    doi = "10.1016/0370-1573(85)90148-6",
    journal = "Phys. Rept.",
    volume = "119",
    pages = "1--74",
    year = "1985"
}

@article{Barvinsky:1987uw,
    author = "Barvinsky, A. O. and Vilkovisky, G. A.",
    title = "{Beyond the Schwinger-Dewitt Technique: Converting Loops Into Trees and In-In Currents}",
    doi = "10.1016/0550-3213(87)90681-X",
    journal = "Nucl. Phys. B",
    volume = "282",
    pages = "163--188",
    year = "1987"
}

@article{Barvinsky:1990up,
    author = "Barvinsky, A. O. and Vilkovisky, G. A.",
    title = "{Covariant perturbation theory. 2: Second order in the curvature. General algorithms}",
    doi = "10.1016/0550-3213(90)90047-H",
    journal = "Nucl. Phys. B",
    volume = "333",
    pages = "471--511",
    year = "1990"
}

@article{Deser:2007jk,
    author = "Deser, Stanley and Woodard, R. P.",
    title = "{Nonlocal Cosmology}",
    eprint = "0706.2151",
    archivePrefix = "arXiv",
    primaryClass = "astro-ph",
    reportNumber = "UFIFT-QG-07-03, BRX-TH-589",
    doi = "10.1103/PhysRevLett.99.111301",
    journal = "Phys. Rev. Lett.",
    volume = "99",
    pages = "111301",
    year = "2007"
}

@article{Woodard:2014iga,
    author = "Woodard, R. P.",
    title = "{Nonlocal Models of Cosmic Acceleration}",
    eprint = "1401.0254",
    archivePrefix = "arXiv",
    primaryClass = "astro-ph.CO",
    reportNumber = "UFIFT-QG-13-09",
    doi = "10.1007/s10701-014-9780-6",
    journal = "Found. Phys.",
    volume = "44",
    pages = "213--233",
    year = "2014"
}

@article{Maggiore:2014sia,
    author = "Maggiore, Michele and Mancarella, Michele",
    title = "{Nonlocal gravity and dark energy}",
    eprint = "1402.0448",
    archivePrefix = "arXiv",
    primaryClass = "hep-th",
    doi = "10.1103/PhysRevD.90.023005",
    journal = "Phys. Rev. D",
    volume = "90",
    number = "2",
    pages = "023005",
    year = "2014"
}

@article{Daas:2024pxs,
    author = "Daas, Jesse and Laporte, Cristobal and Saueressig, Frank and van Dijk, Tim",
    title = "{Rethinking the effective field theory formulation of gravity}",
    eprint = "2405.12685",
    archivePrefix = "arXiv",
    primaryClass = "gr-qc",
    doi = "10.1142/S0218271824410086",
    journal = "Int. J. Mod. Phys. D",
    volume = "33",
    number = "15",
    pages = "2441008",
    year = "2024"
}

@article{Bjerrum-Bohr:2002fji,
    author = "Bjerrum-Bohr, Niels Emil Jannik and Donoghue, John F. and Holstein, Barry R.",
    title = "{Quantum corrections to the Schwarzschild and Kerr metrics}",
    eprint = "hep-th/0211071",
    archivePrefix = "arXiv",
    doi = "10.1103/PhysRevD.68.084005",
    journal = "Phys. Rev. D",
    volume = "68",
    pages = "084005",
    year = "2003",
    note = "[Erratum: Phys.Rev.D 71, 069904 (2005)]"
}

@article{Baez:1999sr,
	author = "Baez, J. C.",
	editor = "Gausterer, H. and Pittner, L. and Grosse, H.",
	title = "{An Introduction to Spin Foam Models of $BF$ Theory and Quantum Gravity}",
	eprint = "gr-qc/9905087",
	archivePrefix = "arXiv",
	doi = "10.1007/3-540-46552-9_2",
	journal = "Lect. Notes Phys.",
	volume = "543",
	pages = "25--93",
	year = "2000"
}

@inproceedings{Perez:2004hj,
	author = "Perez, Alejandro",
	title = "{Introduction to loop quantum gravity and spin foams}",
	booktitle = "{2nd International Conference on Fundamental Interactions}",
	eprint = "gr-qc/0409061",
	archivePrefix = "arXiv",
	month = "9",
	year = "2004"
}

@article{Perez:2012wv,
	author = "Perez, Alejandro",
	title = "{The Spin Foam Approach to Quantum Gravity}",
	eprint = "1205.2019",
	archivePrefix = "arXiv",
	primaryClass = "gr-qc",
	doi = "10.12942/lrr-2013-3",
	journal = "Living Rev. Rel.",
	volume = "16",
	pages = "3",
	year = "2013"
}

@article{Schuller:2009hn,
	author = "Schuller, Frederic P. and Witte, Christof and Wohlfarth, Mattias N. R.",
	title = "{Causal structure and algebraic classification of area metric spacetimes in four dimensions}",
	eprint = "0908.1016",
	archivePrefix = "arXiv",
	primaryClass = "hep-th",
	doi = "10.1016/j.aop.2010.04.008",
	journal = "Annals Phys.",
	volume = "325",
	pages = "1853--1883",
	year = "2010"
}

@article{Borissova:2025frj,
	author = "Borissova, Johanna and Dittrich, Bianca and Eichhorn, Astrid and Schiffer, Marc",
	title = "{Renormalization group flows in area-metric gravity}",
	eprint = "2507.02034",
	archivePrefix = "arXiv",
	primaryClass = "gr-qc",
	month = "7",
	year = "2025"
}

@article{Ho:2015cza,
	author = "Ho, Pei-Ming and Inami, Takeo",
	title = "{Geometry of Area Without Length}",
	eprint = "1508.05569",
	archivePrefix = "arXiv",
	primaryClass = "hep-th",
	doi = "10.1093/ptep/ptv180",
	journal = "PTEP",
	volume = "2016",
	number = "1",
	pages = "013B03",
	year = "2016"
}

@article{Salvio:2018crh,
	author = "Salvio, Alberto",
	title = "{Quadratic Gravity}",
	eprint = "1804.09944",
	archivePrefix = "arXiv",
	primaryClass = "hep-th",
	reportNumber = "CERN-TH-2018-099",
	doi = "10.3389/fphy.2018.00077",
	journal = "Front. in Phys.",
	volume = "6",
	pages = "77",
	year = "2018"
}

@article{Palais:1979rca,
	author = "Palais, Richard S.",
	title = "{The principle of symmetric criticality}",
	doi = "10.1007/BF01941322",
	journal = "Commun. Math. Phys.",
	volume = "69",
	number = "1",
	pages = "19--30",
	year = "1979"
}

@article{Holdom:2002xy,
	author = "Holdom, Bob",
	title = "{On the fate of singularities and horizons in higher derivative gravity}",
	eprint = "hep-th/0206219",
	archivePrefix = "arXiv",
	reportNumber = "UTPT-02-09",
	doi = "10.1103/PhysRevD.66.084010",
	journal = "Phys. Rev. D",
	volume = "66",
	pages = "084010",
	year = "2002"
}

@article{Deser:2003up,
	author = "Deser, Stanley and Tekin, Bayram",
	title = "{Shortcuts to high symmetry solutions in gravitational theories}",
	eprint = "gr-qc/0306114",
	archivePrefix = "arXiv",
	reportNumber = "BRX-TH-520",
	doi = "10.1088/0264-9381/20/22/011",
	journal = "Class. Quant. Grav.",
	volume = "20",
	pages = "4877--4884",
	year = "2003"
}

@article{Schuller:2005yt,
	author = "Schuller, Frederic P. and Wohlfarth, Mattias N. R.",
	title = "{Geometry of manifolds with area metric: multi-metric backgrounds}",
	eprint = "hep-th/0508170",
	archivePrefix = "arXiv",
	reportNumber = "ZMP-HH-05-15",
	doi = "10.1016/j.nuclphysb.2006.04.019",
	journal = "Nucl. Phys. B",
	volume = "747",
	pages = "398--422",
	year = "2006"
}

@article{Schuller:2005ru,
	author = "Schuller, Frederic P. and Wohlfarth, Mattias N. R.",
	title = "{Canonical differential geometry of string backgrounds}",
	eprint = "hep-th/0511157",
	archivePrefix = "arXiv",
	doi = "10.1088/1126-6708/2006/02/059",
	journal = "JHEP",
	volume = "02",
	pages = "059",
	year = "2006"
}

@article{Punzi:2006hy,
	author = "Punzi, Raffaele and Schuller, Frederic P. and Wohlfarth, Mattias N. R.",
	title = "{Geometry for the accelerating universe}",
	eprint = "hep-th/0612133",
	archivePrefix = "arXiv",
	doi = "10.1103/PhysRevD.76.101501",
	journal = "Phys. Rev. D",
	volume = "76",
	pages = "101501",
	year = "2007"
}

@article{Punzi:2006nx,
	author = "Punzi, Raffaele and Schuller, Frederic P. and Wohlfarth, Mattias N. R.",
	title = "{Area metric gravity and accelerating cosmology}",
	eprint = "hep-th/0612141",
	archivePrefix = "arXiv",
	doi = "10.1088/1126-6708/2007/02/030",
	journal = "JHEP",
	volume = "02",
	pages = "030",
	year = "2007"
}

@phdthesis{Perkins:2016imn,
	author = "Perkins, Alun",
	title = "{Static spherically symmetric solutions in higher derivative gravity}",
	doi = "10.25560/44072",
	school = "Imperial Coll., London",
	month = "9",
	year = "2016"
}

@article{Holdom:2016nek,
	author = "Holdom, Bob and Ren, Jing",
	title = "{Not quite a black hole}",
	eprint = "1612.04889",
	archivePrefix = "arXiv",
	primaryClass = "gr-qc",
	doi = "10.1103/PhysRevD.95.084034",
	journal = "Phys. Rev. D",
	volume = "95",
	number = "8",
	pages = "084034",
	year = "2017"
}

@article{Svarc:2018coe,
	author = "\v{S}varc, Robert and Podolsk\'y, Jiri and Pravda, Vojtech and Pravdov\'a, Alena",
	title = "{Exact black holes in quadratic gravity with any cosmological constant}",
	eprint = "1806.09516",
	archivePrefix = "arXiv",
	primaryClass = "gr-qc",
	doi = "10.1103/PhysRevLett.121.231104",
	journal = "Phys. Rev. Lett.",
	volume = "121",
	number = "23",
	pages = "231104",
	year = "2018"
}

@article{Pravda:2016fue,
	author = "Pravda, Vojtech and Pravdov\'a, Alena and Podolsk\'y, Jiri and \v{S}varc, Robert",
	title = "{Exact solutions to quadratic gravity}",
	eprint = "1606.02646",
	archivePrefix = "arXiv",
	primaryClass = "gr-qc",
	doi = "10.1103/PhysRevD.95.084025",
	journal = "Phys. Rev. D",
	volume = "95",
	number = "8",
	pages = "084025",
	year = "2017"
}

@article{Conroy:2014eja,
	author = "Conroy, Aindriu and Koivisto, Tomi and Mazumdar, Anupam and Teimouri, Ali",
	title = "{Generalized quadratic curvature, non-local infrared modifications of gravity and Newtonian potentials}",
	eprint = "1406.4998",
	archivePrefix = "arXiv",
	primaryClass = "hep-th",
	reportNumber = "NORDITA-2014-73",
	doi = "10.1088/0264-9381/32/1/015024",
	journal = "Class. Quant. Grav.",
	volume = "32",
	number = "1",
	pages = "015024",
	year = "2015"
}

@article{Bonanno:2019rsq,
	author = "Bonanno, Alfio and Silveravalle, Samuele",
	title = "{Characterizing black hole metrics in quadratic gravity}",
	eprint = "1903.08759",
	archivePrefix = "arXiv",
	primaryClass = "gr-qc",
	doi = "10.1103/PhysRevD.99.101501",
	journal = "Phys. Rev. D",
	volume = "99",
	number = "10",
	pages = "101501",
	year = "2019"
}

@article{Dittrich:2021kzs,
	author = "Dittrich, Bianca",
	title = "{Modified Graviton Dynamics From Spin Foams: The Area Regge Action}",
	eprint = "2105.10808",
	archivePrefix = "arXiv",
	primaryClass = "gr-qc",
	month = "5",
	year = "2021"
}

@article{Giacchini:2021pmr,
	author = "Giacchini, Breno L. and Netto, Tib\'erio de Paula and Modesto, Leonardo",
	title = "{Action principle selection of regular black holes}",
	eprint = "2105.00300",
	archivePrefix = "arXiv",
	primaryClass = "gr-qc",
	doi = "10.1103/PhysRevD.104.084072",
	journal = "Phys. Rev. D",
	volume = "104",
	number = "8",
	pages = "084072",
	year = "2021"
}

@article{Dittrich:2022yoo,
	author = "Dittrich, Bianca and Kogios, Athanasios",
	title = "{From spin foams to area metric dynamics to gravitons}",
	eprint = "2203.02409",
	archivePrefix = "arXiv",
	primaryClass = "gr-qc",
	doi = "10.1088/1361-6382/acc5d9",
	journal = "Class. Quant. Grav.",
	volume = "40",
	number = "9",
	pages = "095011",
	year = "2023"
}

@article{Bonanno:2022ibv,
	author = "Bonanno, Alfio and Silveravalle, Samuele and Zuccotti, Alessandro",
	title = "{Nonsymmetric wormholes and localized big rip singularities in Einstein-Weyl gravity}",
	eprint = "2204.04966",
	archivePrefix = "arXiv",
	primaryClass = "gr-qc",
	doi = "10.1103/PhysRevD.105.124059",
	journal = "Phys. Rev. D",
	volume = "105",
	number = "12",
	pages = "124059",
	year = "2022"
}

@article{Borissova:2022clg,
	author = "Borissova, Johanna N. and Dittrich, Bianca",
	title = "{Towards effective actions for the continuum limit of spin foams}",
	eprint = "2207.03307",
	archivePrefix = "arXiv",
	primaryClass = "gr-qc",
	doi = "10.1088/1361-6382/accbfb",
	journal = "Class. Quant. Grav.",
	volume = "40",
	number = "10",
	pages = "105006",
	year = "2023"
}

@article{Borissova:2023kzq,
	author = "Borissova, Johanna N.",
	title = "{Suppression of spacetime singularities in quantum gravity}",
	eprint = "2309.05695",
	archivePrefix = "arXiv",
	primaryClass = "gr-qc",
	doi = "10.1088/1361-6382/ad46c0",
	journal = "Class. Quant. Grav.",
	volume = "41",
	number = "12",
	pages = "127002",
	year = "2024"
}

@article{Dittrich:2023ava,
	author = {Dittrich, Bianca and Padua-Arg\"uelles, Jos\'e},
	title = "{Twisted geometries are area-metric geometries}",
	eprint = "2302.11586",
	archivePrefix = "arXiv",
	primaryClass = "gr-qc",
	doi = "10.1103/PhysRevD.109.026002",
	journal = "Phys. Rev. D",
	volume = "109",
	number = "2",
	pages = "026002",
	year = "2024"
}

@article{Borissova:2023yxs,
	author = "Borissova, Johanna N. and Dittrich, Bianca and Krasnov, Kirill",
	title = "{Area-metric gravity revisited}",
	eprint = "2312.13935",
	archivePrefix = "arXiv",
	primaryClass = "gr-qc",
	doi = "10.1103/PhysRevD.109.124035",
	journal = "Phys. Rev. D",
	volume = "109",
	number = "12",
	pages = "124035",
	year = "2024"
}

@article{Borissova:2024cpx,
	author = "Borissova, Johanna and Ho, Pei-Ming",
	title = "{From area metric backgrounds to the cosmological constant and corrections to the Polyakov action}",
	eprint = "2404.14478",
	archivePrefix = "arXiv",
	primaryClass = "hep-th",
	doi = "10.1103/PhysRevD.110.046017",
	journal = "Phys. Rev. D",
	volume = "110",
	number = "4",
	pages = "046017",
	year = "2024"
}

@inproceedings{Buoninfante:2024oxl,
	author = "Afshordi, Niayesh and others",
	editor = "Buoninfante, Luca and Carballo-Rubio, Ra\'ul and Cardoso, Vitor and Di Filippo, Francesco and Eichhorn, Astrid",
	title = "{Black Holes Inside and Out 2024: visions for the future of black hole physics}",
	eprint = "2410.14414",
	archivePrefix = "arXiv",
	primaryClass = "gr-qc",
	month = "10",
	year = "2024"
}

@article{Carballo-Rubio:2025fnc,
	author = "Carballo-Rubio, Ra\'ul and others",
	title = "{Towards a Non-singular Paradigm of Black Hole Physics}",
	eprint = "2501.05505",
	archivePrefix = "arXiv",
	primaryClass = "gr-qc",
	month = "1",
	year = "2025"
}

@article{Giacchini:2024exc,
	author = "Giacchini, Breno L. and Kol{\'a}{\v{r}}, Ivan",
	title = "{Toward regular black holes in sixth-derivative gravity}",
	eprint = "2406.00997",
	archivePrefix = "arXiv",
	primaryClass = "gr-qc",
	doi = "10.1103/PhysRevD.110.104056",
	journal = "Phys. Rev. D",
	volume = "110",
	number = "10",
	pages = "104056",
	year = "2024"
}

@article{Frausto:2024egp,
    author = "Frausto, Guillermo and Kol{\'a}{\v{r}}, Ivan and M{\'a}lek, Tom{\'a}{\v{s}} and Torre, Charles",
    title = "{Symmetry reduction of gravitational Lagrangians}",
    eprint = "2410.11036",
    archivePrefix = "arXiv",
    primaryClass = "gr-qc",
    doi = "10.1103/PhysRevD.111.064062",
    journal = "Phys. Rev. D",
    volume = "111",
    number = "6",
    pages = "064062",
    year = "2025"
}

@article{deRham:2020ejn,
    author = "de Rham, Claudia and Francfort, J{\'e}r{\'e}mie and Zhang, Jun",
    title = "{Black Hole Gravitational Waves in the Effective Field Theory of Gravity}",
    eprint = "2005.13923",
    archivePrefix = "arXiv",
    primaryClass = "hep-th",
    reportNumber = "Imperial/TP/2020/CdR/02",
    doi = "10.1103/PhysRevD.102.024079",
    journal = "Phys. Rev. D",
    volume = "102",
    number = "2",
    pages = "024079",
    year = "2020"
}

@article{Donoghue:1994dn,
    author = "Donoghue, John F.",
    title = "{General relativity as an effective field theory: The leading quantum corrections}",
    eprint = "gr-qc/9405057",
    archivePrefix = "arXiv",
    reportNumber = "UMHEP-408",
    doi = "10.1103/PhysRevD.50.3874",
    journal = "Phys. Rev. D",
    volume = "50",
    pages = "3874--3888",
    year = "1994"
}

@article{Stelle:1977ry,
    author = "Stelle, K. S.",
    title = "{Classical Gravity with Higher Derivatives}",
    reportNumber = "Print-77-0417 (BRANDEIS)",
    doi = "10.1007/BF00760427",
    journal = "Gen. Rel. Grav.",
    volume = "9",
    pages = "353--371",
    year = "1978"
}

@article{Lu:2015cqa,
    author = "L{\"u}, H. and Perkins, A. and Pope, C. N. and Stelle, K. S.",
    title = "{Black Holes in Higher-Derivative Gravity}",
    eprint = "1502.01028",
    archivePrefix = "arXiv",
    primaryClass = "hep-th",
    reportNumber = "IMPERIAL-TP-15-KSS-01, MI-TH-1504, CAQS-1501",
    doi = "10.1103/PhysRevLett.114.171601",
    journal = "Phys. Rev. Lett.",
    volume = "114",
    number = "17",
    pages = "171601",
    year = "2015"
}

@article{Lu:2015psa,
    author = {L{\"u}, H. and Perkins, A. and Pope, C. N. and Stelle, K. S.},
    title = "{Spherically Symmetric Solutions in Higher-Derivative Gravity}",
    eprint = "1508.00010",
    archivePrefix = "arXiv",
    primaryClass = "hep-th",
    reportNumber = "IMPERIAL-TP-15-KSS-02, MI-TH-1528",
    doi = "10.1103/PhysRevD.92.124019",
    journal = "Phys. Rev. D",
    volume = "92",
    number = "12",
    pages = "124019",
    year = "2015"
}

@article{Podolsky:2018pfe,
    author = "Podolsk\'y, Jiri and \v{S}varc, Robert and Pravda, Vojtech and Pravdov\'a, Alena",
    title = "{Explicit black hole solutions in higher-derivative gravity}",
    eprint = "1806.08209",
    archivePrefix = "arXiv",
    primaryClass = "gr-qc",
    doi = "10.1103/PhysRevD.98.021502",
    journal = "Phys. Rev. D",
    volume = "98",
    number = "2",
    pages = "021502",
    year = "2018"
}

@article{Podolsky:2019gro,
    author = "Podolsk{\'y}, Jiri and {\v{S}}varc, Robert and Pravda, Vojtech and Pravdov\'a, Alena",
    title = "{Black holes and other exact spherical solutions in Quadratic Gravity}",
    eprint = "1907.00046",
    archivePrefix = "arXiv",
    primaryClass = "gr-qc",
    doi = "10.1103/PhysRevD.101.024027",
    journal = "Phys. Rev. D",
    volume = "101",
    number = "2",
    pages = "024027",
    year = "2020"
}

@article{Held:2022abx,
    author = "Held, Aaron and Zhang, Jun",
    title = "{Instability of spherically symmetric black holes in quadratic gravity}",
    eprint = "2209.01867",
    archivePrefix = "arXiv",
    primaryClass = "gr-qc",
    reportNumber = "Imperial/TP/2022/AH/03",
    doi = "10.1103/PhysRevD.107.064060",
    journal = "Phys. Rev. D",
    volume = "107",
    number = "6",
    pages = "064060",
    year = "2023"
}

@article{Giacchini:2025mlv,
    author = "Giacchini, Breno L. and Kol{\'a}{\v{r}}, Ivan",
    title = "{Neglected solutions in quadratic gravity}",
    eprint = "2509.07317",
    archivePrefix = "arXiv",
    primaryClass = "gr-qc",
    doi = "10.1103/zptz-srkl",
    journal = "Phys. Rev. D",
    volume = "113",
    number = "4",
    pages = "L041502",
    year = "2026"
}

@article{Giacchini:2025gzw,
    author = "Giacchini, Breno L. and Kol{\'a}{\v{r}}, Ivan",
    title = "{Black holes and other exact solutions in six-derivative gravity}",
    eprint = "2503.17318",
    archivePrefix = "arXiv",
    primaryClass = "gr-qc",
    doi = "10.1103/1mkn-kfpx",
    journal = "Phys. Rev. D",
    volume = "112",
    number = "2",
    pages = "024051",
    year = "2025"
}

@article{Cardoso:2018ptl,
    author = "Cardoso, Vitor and Kimura, Masashi and Maselli, Andrea and Senatore, Leonardo",
    title = "{Black Holes in an Effective Field Theory Extension of General Relativity}",
    eprint = "1808.08962",
    archivePrefix = "arXiv",
    primaryClass = "gr-qc",
    doi = "10.1103/PhysRevLett.121.251105",
    journal = "Phys. Rev. Lett.",
    volume = "121",
    number = "25",
    pages = "251105",
    year = "2018",
    note = "[Erratum: Phys.Rev.Lett. 131, 109903 (2023)]"
}

@article{Cano:2019ore,
    author = "Cano, Pablo A. and Ruip{\'e}rez, Alejandro",
    title = "{Leading higher-derivative corrections to Kerr geometry}",
    eprint = "1901.01315",
    archivePrefix = "arXiv",
    primaryClass = "gr-qc",
    reportNumber = "IFT-UAM/CSIC-19-2",
    doi = "10.1007/JHEP05(2019)189",
    journal = "JHEP",
    volume = "05",
    pages = "189",
    year = "2019",
    note = "[Erratum: JHEP 03, 187 (2020)]"
}

@article{Cano:2020cao,
    author = "Cano, Pablo A. and Fransen, Kwinten and Hertog, Thomas",
    title = "{Ringing of rotating black holes in higher-derivative gravity}",
    eprint = "2005.03671",
    archivePrefix = "arXiv",
    primaryClass = "gr-qc",
    doi = "10.1103/PhysRevD.102.044047",
    journal = "Phys. Rev. D",
    volume = "102",
    number = "4",
    pages = "044047",
    year = "2020"
}

@article{Endlich:2017tqa,
    author = "Endlich, Solomon and Gorbenko, Victor and Huang, Junwu and Senatore, Leonardo",
    title = "{An effective formalism for testing extensions to General Relativity with gravitational waves}",
    eprint = "1704.01590",
    archivePrefix = "arXiv",
    primaryClass = "gr-qc",
    doi = "10.1007/JHEP09(2017)122",
    journal = "JHEP",
    volume = "09",
    pages = "122",
    year = "2017"
}

@article{Sennett:2019bpc,
    author = "Sennett, Noah and Brito, Richard and Buonanno, Alessandra and Gorbenko, Victor and Senatore, Leonardo",
    title = "{Gravitational-Wave Constraints on an Effective Field-Theory Extension of General Relativity}",
    eprint = "1912.09917",
    archivePrefix = "arXiv",
    primaryClass = "gr-qc",
    doi = "10.1103/PhysRevD.102.044056",
    journal = "Phys. Rev. D",
    volume = "102",
    number = "4",
    pages = "044056",
    year = "2020"
}

@article{Bosma:2019aiu,
    author = "Bosma, Lando and Knorr, Benjamin and Saueressig, Frank",
    title = "{Resolving Spacetime Singularities within Asymptotic Safety}",
    eprint = "1904.04845",
    archivePrefix = "arXiv",
    primaryClass = "hep-th",
    doi = "10.1103/PhysRevLett.123.101301",
    journal = "Phys. Rev. Lett.",
    volume = "123",
    number = "10",
    pages = "101301",
    year = "2019"
}

@article{Knorr:2018kog,
    author = "Knorr, Benjamin and Saueressig, Frank",
    title = "{Towards reconstructing the quantum effective action of gravity}",
    eprint = "1804.03846",
    archivePrefix = "arXiv",
    primaryClass = "hep-th",
    doi = "10.1103/PhysRevLett.121.161304",
    journal = "Phys. Rev. Lett.",
    volume = "121",
    number = "16",
    pages = "161304",
    year = "2018"
}

@article{Calcagni:2014vxa,
    author = "Calcagni, Gianluca and Modesto, Leonardo",
    title = "{Nonlocal quantum gravity and M-theory}",
    eprint = "1404.2137",
    archivePrefix = "arXiv",
    primaryClass = "hep-th",
    doi = "10.1103/PhysRevD.91.124059",
    journal = "Phys. Rev. D",
    volume = "91",
    number = "12",
    pages = "124059",
    year = "2015"
}

@article{Knorr:2021niv,
    author = "Knorr, Benjamin and Schiffer, Marc",
    title = "{Non-Perturbative Propagators in Quantum Gravity}",
    eprint = "2105.04566",
    archivePrefix = "arXiv",
    primaryClass = "hep-th",
    doi = "10.3390/universe7070216",
    journal = "Universe",
    volume = "7",
    number = "7",
    pages = "216",
    year = "2021"
}

@article{Fehre:2021eob,
    author = "Fehre, Jannik and Litim, Daniel F. and Pawlowski, Jan M. and Reichert, Manuel",
    title = "{Lorentzian Quantum Gravity and the Graviton Spectral Function}",
    eprint = "2111.13232",
    archivePrefix = "arXiv",
    primaryClass = "hep-th",
    doi = "10.1103/PhysRevLett.130.081501",
    journal = "Phys. Rev. Lett.",
    volume = "130",
    number = "8",
    pages = "081501",
    year = "2023"
}

@article{Fradkin:1985ys,
    author = "Fradkin, E. S. and Tseytlin, Arkady A.",
    title = "{Quantum String Theory Effective Action}",
    doi = "10.1016/0550-3213(85)90559-0",
    journal = "Nucl. Phys. B",
    volume = "261",
    pages = "1--27",
    year = "1985",
    note = "[Erratum: Nucl.Phys.B 269, 745--745 (1986)]"
}

@article{Pawlowski:2023dda,
    author = {Pawlowski, Jan M. and Tr{\"a}nkle, Jan},
    title = "{Effective action and black hole solutions in asymptotically safe quantum gravity}",
    eprint = "2309.17043",
    archivePrefix = "arXiv",
    primaryClass = "hep-th",
    doi = "10.1103/PhysRevD.110.086011",
    journal = "Phys. Rev. D",
    volume = "110",
    number = "8",
    pages = "086011",
    year = "2024"
}

@article{Calcagni:2013eua,
    author = "Calcagni, Gianluca and Modesto, Leonardo",
    title = "{Nonlocality in string theory}",
    eprint = "1310.4957",
    archivePrefix = "arXiv",
    primaryClass = "hep-th",
    doi = "10.1088/1751-8113/47/35/355402",
    journal = "J. Phys. A",
    volume = "47",
    number = "35",
    pages = "355402",
    year = "2014"
}

@article{Zwiebach:1992ie,
    author = "Zwiebach, Barton",
    title = "{Closed string field theory: Quantum action and the B-V master equation}",
    eprint = "hep-th/9206084",
    archivePrefix = "arXiv",
    reportNumber = "IASSNS-HEP-92-41, MIT-CTP-2102",
    doi = "10.1016/0550-3213(93)90388-6",
    journal = "Nucl. Phys. B",
    volume = "390",
    pages = "33--152",
    year = "1993"
}

@article{Pawlowski:2025etp,
    author = "Pawlowski, Jan M. and Reichert, Manuel and Wessely, Jonas",
    title = "{Self-consistent graviton spectral function in Lorentzian quantum gravity}",
    eprint = "2507.22169",
    archivePrefix = "arXiv",
    primaryClass = "hep-th",
    month = "7",
    year = "2025"
}

@article{Bonanno:2021squ,
    author = "Bonanno, Alfio and Denz, Tobias and Pawlowski, Jan M. and Reichert, Manuel",
    title = "{Reconstructing the graviton}",
    eprint = "2102.02217",
    archivePrefix = "arXiv",
    primaryClass = "hep-th",
    doi = "10.21468/SciPostPhys.12.1.001",
    journal = "SciPost Phys.",
    volume = "12",
    number = "1",
    pages = "001",
    year = "2022"
}

@article{Knorr:2019atm,
    author = "Knorr, Benjamin and Ripken, Chris and Saueressig, Frank",
    title = "{Form Factors in Asymptotic Safety: conceptual ideas and computational toolbox}",
    eprint = "1907.02903",
    archivePrefix = "arXiv",
    primaryClass = "hep-th",
    doi = "10.1088/1361-6382/ab4a53",
    journal = "Class. Quant. Grav.",
    volume = "36",
    number = "23",
    pages = "234001",
    year = "2019"
}

@article{Christiansen:2014raa,
    author = "Christiansen, Nicolai and Knorr, Benjamin and Pawlowski, Jan M. and Rodigast, Andreas",
    title = "{Global Flows in Quantum Gravity}",
    eprint = "1403.1232",
    archivePrefix = "arXiv",
    primaryClass = "hep-th",
    doi = "10.1103/PhysRevD.93.044036",
    journal = "Phys. Rev. D",
    volume = "93",
    number = "4",
    pages = "044036",
    year = "2016"
}

@article{Bernard:2025dyh,
    author = "Bernard, Laura and Giri, Suvendu and Lehner, Luis and Sturani, Riccardo",
    title = "{Generic EFT-motivated beyond General Relativity gravitational wave tests and their curvature dependence: from observation to interpretation}",
    eprint = "2507.17143",
    archivePrefix = "arXiv",
    primaryClass = "gr-qc",
    reportNumber = "UUITP-20/25",
    month = "7",
    year = "2025"
}

@article{Held:2021pht,
    author = "Held, Aaron and Lim, Hyun",
    title = "{Nonlinear dynamics of quadratic gravity in spherical symmetry}",
    eprint = "2104.04010",
    archivePrefix = "arXiv",
    primaryClass = "gr-qc",
    reportNumber = "LA-UR-21-22739, Imperial/TP/2021/AH/03",
    doi = "10.1103/PhysRevD.104.084075",
    journal = "Phys. Rev. D",
    volume = "104",
    number = "8",
    pages = "084075",
    year = "2021"
}

@article{Cayuso:2020lca,
    author = "Cayuso, Ramiro and Lehner, Luis",
    title = "{Nonlinear, noniterative treatment of EFT-motivated gravity}",
    eprint = "2005.13720",
    archivePrefix = "arXiv",
    primaryClass = "gr-qc",
    doi = "10.1103/PhysRevD.102.084008",
    journal = "Phys. Rev. D",
    volume = "102",
    number = "8",
    pages = "084008",
    year = "2020"
}

@article{Held:2023aap,
    author = "Held, Aaron and Lim, Hyun",
    title = "{Nonlinear evolution of quadratic gravity in 3+1 dimensions}",
    eprint = "2306.04725",
    archivePrefix = "arXiv",
    primaryClass = "gr-qc",
    reportNumber = "LA-UR-23-23440",
    doi = "10.1103/PhysRevD.108.104025",
    journal = "Phys. Rev. D",
    volume = "108",
    number = "10",
    pages = "104025",
    year = "2023"
}

@article{Cayuso:2023aht,
    author = "Cayuso, Ramiro and Figueras, Pau and Fran{\c{c}}a, Tiago and Lehner, Luis",
    title = "{Self-Consistent Modeling of Gravitational Theories beyond General Relativity}",
    eprint = "2303.07246",
    archivePrefix = "arXiv",
    primaryClass = "gr-qc",
    doi = "10.1103/PhysRevLett.131.111403",
    journal = "Phys. Rev. Lett.",
    volume = "131",
    number = "11",
    pages = "111403",
    year = "2023"
}

@article{Figueras:2024bba,
    author = "Figueras, Pau and Held, Aaron and Kov{\'a}cs, {\'A}ron D.",
    title = "{Well-posed initial value formulation of general effective field theories of gravity}",
    eprint = "2407.08775",
    archivePrefix = "arXiv",
    primaryClass = "gr-qc",
    month = "7",
    year = "2024"
}

@article{Held:2025ckb,
    author = "Held, Aaron and Lim, Hyun",
    title = "{Black-hole binaries and waveforms in Quadratic Gravity}",
    eprint = "2503.13428",
    archivePrefix = "arXiv",
    primaryClass = "gr-qc",
    reportNumber = "LA-UR-24-24999",
    month = "3",
    year = "2025"
}

@article{Buoninfante:2024yth,
	author = "Buoninfante, Luca and others",
	title = "{Visions in quantum gravity}",
	eprint = "2412.08696",
	archivePrefix = "arXiv",
	primaryClass = "hep-th",
	doi = "10.21468/SciPostPhysCommRep.11",
	month = "12",
	year = "2024"
}

@inproceedings{Bambi:2025wjx,
	author = "Bambi, Cosimo and others",
	title = "{Black hole mimickers: from theory to observation}",
	eprint = "2505.09014",
	archivePrefix = "arXiv",
	primaryClass = "gr-qc",
	month = "5",
	year = "2025"
}

@article{Kokkotas:2017zwt,
    author = "Kokkotas, K. and Konoplya, R. A. and Zhidenko, A.",
    title = "{Non-Schwarzschild black-hole metric in four dimensional higher derivative gravity: analytical approximation}",
    eprint = "1705.09875",
    archivePrefix = "arXiv",
    primaryClass = "gr-qc",
    doi = "10.1103/PhysRevD.96.064007",
    journal = "Phys. Rev. D",
    volume = "96",
    pages = "064007",
    year = "2017"
}

@article{Surya:2019ndm,
	author = "Surya, Sumati",
	title = "{The causal set approach to quantum gravity}",
	eprint = "1903.11544",
	archivePrefix = "arXiv",
	primaryClass = "gr-qc",
	doi = "10.1007/s41114-019-0023-1",
	journal = "Living Rev. Rel.",
	volume = "22",
	number = "1",
	pages = "5",
	year = "2019"
}

@article{Henson:2006kf,
	author = "Henson, Joe",
	title = "{The Causal set approach to quantum gravity}",
	eprint = "gr-qc/0601121",
	archivePrefix = "arXiv",
	pages = "393--413",
	month = "1",
	year = "2006"
}

@article{Rovelli:1997yv,
	author = "Rovelli, Carlo",
	title = "{Loop quantum gravity}",
	eprint = "gr-qc/9710008",
	archivePrefix = "arXiv",
	doi = "10.12942/lrr-1998-1",
	journal = "Living Rev. Rel.",
	volume = "1",
	pages = "1",
	year = "1998"
}

@article{Ashtekar:2021kfp,
	author = "Ashtekar, Abhay and Bianchi, Eugenio",
	title = "{A short review of loop quantum gravity}",
	eprint = "2104.04394",
	archivePrefix = "arXiv",
	primaryClass = "gr-qc",
	doi = "10.1088/1361-6633/abed91",
	journal = "Rept. Prog. Phys.",
	volume = "84",
	number = "4",
	pages = "042001",
	year = "2021"
}

@inproceedings{Sahlmann:2010zf,
	author = "Sahlmann, Hanno",
	title = "{Loop Quantum Gravity - A Short Review}",
	booktitle = "{Foundations of Space and Time: Reflections on Quantum Gravity}",
	eprint = "1001.4188",
	archivePrefix = "arXiv",
	primaryClass = "gr-qc",
	reportNumber = "KA-TP-19-2009",
	pages = "185--210",
	month = "1",
	year = "2010"
}

@article{Borissova:2020knn,
	author = "Borissova, Johanna N. and Eichhorn, Astrid",
	title = "{Towards black-hole singularity-resolution in the Lorentzian gravitational path integral}",
	eprint = "2012.08570",
	archivePrefix = "arXiv",
	primaryClass = "gr-qc",
	doi = "10.3390/universe7030048",
	journal = "Universe",
	volume = "7",
	number = "3",
	pages = "48",
	year = "2021"
}

@article{Borissova:2024hkc,
	author = "Borissova, Johanna and Eichhorn, Astrid and Ray, Shouryya",
	title = "{A non-local way around the no-global-symmetries conjecture in quantum gravity?}",
	eprint = "2407.09595",
	archivePrefix = "arXiv",
	primaryClass = "hep-th",
	doi = "10.1088/1361-6382/ada2d4",
	journal = "Class. Quant. Grav.",
	volume = "42",
	number = "3",
	pages = "037001",
	year = "2025"
}

@article{DelPorro:2025wts,
	author = "Del Porro, Francesco and Pfeiffer, Jonas and Platania, Alessia and Silveravalle, Samuele",
	title = "{Charting GLOBs in Asymptotically Safe Gravity}",
	eprint = "2509.14309",
	archivePrefix = "arXiv",
	primaryClass = "gr-qc",
	month = "9",
	year = "2025"
}

@article{Silveravalle:2022wij,
	author = "Silveravalle, Samuele and Zuccotti, Alessandro",
	title = "{Phase diagram of Einstein-Weyl gravity}",
	eprint = "2210.13877",
	archivePrefix = "arXiv",
	primaryClass = "gr-qc",
	doi = "10.1103/PhysRevD.107.064029",
	journal = "Phys. Rev. D",
	volume = "107",
	number = "6",
	pages = "064029",
	year = "2023"
}

@article{Maggiore:2013mea,
	author = "Maggiore, Michele",
	title = "{Phantom dark energy from nonlocal infrared modifications of general relativity}",
	eprint = "1307.3898",
	archivePrefix = "arXiv",
	primaryClass = "hep-th",
	doi = "10.1103/PhysRevD.89.043008",
	journal = "Phys. Rev. D",
	volume = "89",
	number = "4",
	pages = "043008",
	year = "2014"
}

@article{Foffa:2013vma,
	author = "Foffa, Stefano and Maggiore, Michele and Mitsou, Ermis",
	title = "{Cosmological dynamics and dark energy from nonlocal infrared modifications of gravity}",
	eprint = "1311.3435",
	archivePrefix = "arXiv",
	primaryClass = "hep-th",
	doi = "10.1142/S0217751X14501164",
	journal = "Int. J. Mod. Phys. A",
	volume = "29",
	pages = "1450116",
	year = "2014"
}

@article{Ferreira:2013tqn,
    author = "Ferreira, Pedro G. and Maroto, Antonio L.",
    title = "{A few cosmological implications of tensor nonlocalities}",
    eprint = "1310.1238",
    archivePrefix = "arXiv",
    primaryClass = "astro-ph.CO",
    doi = "10.1103/PhysRevD.88.123502",
    journal = "Phys. Rev. D",
    volume = "88",
    number = "12",
    pages = "123502",
    year = "2013"
}

@Inbook{Donoghue:2022eay,
author="Donoghue, John F.",
editor="Bambi, Cosimo
and Modesto, Leonardo
and Shapiro, Ilya",
title="Quantum General Relativity and Effective Field Theory",
bookTitle="Handbook of Quantum Gravity",
year="2024",
publisher="Springer Nature Singapore",
address="Singapore",
pages="3--26",
isbn="978-981-99-7681-2",
doi="10.1007/978-981-99-7681-2_1",
url="https://doi.org/10.1007/978-981-99-7681-2_1"
}

@article{Gorbar:2003yt,
    author = "Gorbar, Eduard V. and Shapiro, Ilya L.",
    title = "{Renormalization group and decoupling in curved space. 2. The Standard model and beyond}",
    eprint = "hep-ph/0303124",
    archivePrefix = "arXiv",
    reportNumber = "DF-UFJF-03-01",
    doi = "10.1088/1126-6708/2003/06/004",
    journal = "JHEP",
    volume = "06",
    pages = "004",
    year = "2003"
}

@article{Gorbar:2002pw,
    author = "Gorbar, E. V. and Shapiro, I. L.",
    title = "{Renormalization group and decoupling in curved space}",
    eprint = "hep-ph/0210388",
    archivePrefix = "arXiv",
    reportNumber = "DF-UFJF-02-02",
    doi = "10.1088/1126-6708/2003/02/021",
    journal = "JHEP",
    volume = "02",
    pages = "021",
    year = "2003"
}

@article{Gorbar:2003yp,
    author = "Gorbar, Eduard V. and Shapiro, Ilya L.",
    title = "{Renormalization group and decoupling in curved space. 3. The Case of spontaneous symmetry breaking}",
    eprint = "hep-ph/0311190",
    archivePrefix = "arXiv",
    reportNumber = "DF-UFJF-03-09",
    doi = "10.1088/1126-6708/2004/02/060",
    journal = "JHEP",
    volume = "02",
    pages = "060",
    year = "2004"
}

@article{Barvinsky:2003kg,
    author = "Barvinsky, A. O.",
    title = "{Nonlocal action for long distance modifications of gravity theory}",
    eprint = "hep-th/0304229",
    archivePrefix = "arXiv",
    doi = "10.1016/j.physletb.2003.08.055",
    journal = "Phys. Lett. B",
    volume = "572",
    pages = "109--116",
    year = "2003"
}

@article{Knorr:2022kqp,
	author = "Knorr, Benjamin and Platania, Alessia",
	title = "{Sifting quantum black holes through the principle of least action}",
	eprint = "2202.01216",
	archivePrefix = "arXiv",
	primaryClass = "hep-th",
	doi = "10.1103/PhysRevD.106.L021901",
	journal = "Phys. Rev. D",
	volume = "106",
	number = "2",
	pages = "L021901",
	year = "2022"
}

@article{Stelle:1976gc,
    author = "Stelle, K. S.",
    title = "{Renormalization of Higher Derivative Quantum Gravity}",
    reportNumber = "PRINT-76-1059 (BRANDEIS)",
    doi = "10.1103/PhysRevD.16.953",
    journal = "Phys. Rev. D",
    volume = "16",
    pages = "953--969",
    year = "1977"
}

@book{Buchbinder:2021wzv,
    author = "Buchbinder, Iosif L. and Shapiro, Ilya",
    title = "{Introduction to Quantum Field Theory with Applications to Quantum Gravity}",
    doi = "10.1093/oso/9780198838319.001.0001",
    isbn = "978-0-19-887234-4, 978-0-19-883831-9",
    publisher = "Oxford University Press",
    series = "Oxford Graduate Texts",
    month = "3",
    year = "2021"
}

@article{tHooft:1974toh,
    author = "'t Hooft, Gerard and Veltman, M. J. G.",
    title = "{One-loop divergencies in the theory of gravitation}",
    doi = "10.1142/9789814539395_0001",
    journal = "Ann. Inst. H. Poincare Phys. Theor. A",
    volume = "20",
    number = "1",
    pages = "69--94",
    year = "1974"
}

@article{Deser:1974cz,
    author = "Deser, Stanley and van Nieuwenhuizen, P.",
    title = "{One Loop Divergences of Quantized Einstein-Maxwell Fields}",
    reportNumber = "Print-74-0576 (BRANDEIS)",
    doi = "10.1103/PhysRevD.10.401",
    journal = "Phys. Rev. D",
    volume = "10",
    pages = "401",
    year = "1974"
}

@article{Deser:1974xq,
    author = "Deser, Stanley and Tsao, Hung-Sheng and van Nieuwenhuizen, P.",
    title = "{One Loop Divergences of the Einstein Yang-Mills System}",
    reportNumber = "Print-74-1164 (BRANDEIS)",
    doi = "10.1103/PhysRevD.10.3337",
    journal = "Phys. Rev. D",
    volume = "10",
    pages = "3337",
    year = "1974"
}

@article{Goroff:1985th,
    author = "Goroff, Marc H. and Sagnotti, Augusto",
    title = "{The Ultraviolet Behavior of Einstein Gravity}",
    reportNumber = "CALT-68-1289, LBL-19995, UCB-PTH-85-34",
    doi = "10.1016/0550-3213(86)90193-8",
    journal = "Nucl. Phys. B",
    volume = "266",
    pages = "709--736",
    year = "1986"
}

@article{Appelquist:1974tg,
    author = "Appelquist, Thomas and Carazzone, J.",
    title = "{Infrared Singularities and Massive Fields}",
    reportNumber = "Print-74-1486 (HARVARD)",
    doi = "10.1103/PhysRevD.11.2856",
    journal = "Phys. Rev. D",
    volume = "11",
    pages = "2856",
    year = "1975"
}

@book{Hawking:1973uf,
	author = "Hawking, Stephen W. and Ellis, George F. R.",
	title = "{The Large Scale Structure of Space-Time}",
	doi = "10.1017/9781009253161",
	isbn = "978-1-009-25316-1, 978-1-009-25315-4, 978-0-521-20016-5, 978-0-521-09906-6, 978-0-511-82630-6, 978-0-521-09906-6",
	publisher = "Cambridge University Press",
	series = "Cambridge Monographs on Mathematical Physics",
	month = "2",
	year = "2023"
}

@article{Penrose:1964wq,
	author = "Penrose, Roger",
	title = "{Gravitational collapse and space-time singularities}",
	doi = "10.1103/PhysRevLett.14.57",
	journal = "Phys. Rev. Lett.",
	volume = "14",
	pages = "57--59",
	year = "1965"
}

@article{Asorey:1996hz,
    author = "Asorey, M. and L\'opez, J. L. and Shapiro, I. L.",
    title = "{Some remarks on high derivative quantum gravity}",
    eprint = "hep-th/9610006",
    archivePrefix = "arXiv",
    reportNumber = "DFTUZ-96-15",
    doi = "10.1142/S0217751X97002991",
    journal = "Int. J. Mod. Phys. A",
    volume = "12",
    pages = "5711--5734",
    year = "1997"
}

@article{Bhattacharya:2025hag,
    author = "Bhattacharya, Aranya and Chawla, Lavish and Flory, Mario and Kulig, Mateusz",
    title = "{On the role of Area Metrics in AdS/CFT}",
    eprint = "2511.20753",
    archivePrefix = "arXiv",
    primaryClass = "hep-th",
    month = "11",
    year = "2025"
}

@article{Kovacs:2020pns,
    author = "Kov{\'a}cs, {\'A}ron D. and Reall, Harvey S.",
    title = "{Well-Posed Formulation of Scalar-Tensor Effective Field Theory}",
    eprint = "2003.04327",
    archivePrefix = "arXiv",
    primaryClass = "gr-qc",
    doi = "10.1103/PhysRevLett.124.221101",
    journal = "Phys. Rev. Lett.",
    volume = "124",
    number = "22",
    pages = "221101",
    year = "2020"
}

@article{Kovacs:2020ywu,
    author = "Kov{\'a}cs, {\'A}ron D. and Reall, Harvey S.",
    title = "{Well-posed formulation of Lovelock and Horndeski theories}",
    eprint = "2003.08398",
    archivePrefix = "arXiv",
    primaryClass = "gr-qc",
    doi = "10.1103/PhysRevD.101.124003",
    journal = "Phys. Rev. D",
    volume = "101",
    number = "12",
    pages = "124003",
    year = "2020"
}

@article{Ripley:2019aqj,
    author = "Ripley, Justin L. and Pretorius, Frans",
    title = "{Scalarized Black Hole dynamics in Einstein dilaton Gauss-Bonnet Gravity}",
    eprint = "1911.11027",
    archivePrefix = "arXiv",
    primaryClass = "gr-qc",
    doi = "10.1103/PhysRevD.101.044015",
    journal = "Phys. Rev. D",
    volume = "101",
    number = "4",
    pages = "044015",
    year = "2020"
}

@article{Ripley:2020vpk,
    author = "Ripley, Justin L. and Pretorius, Frans",
    title = "{Dynamics of a $\mathbb Z_2$ symmetric EdGB gravity in spherical symmetry}",
    eprint = "2005.05417",
    archivePrefix = "arXiv",
    primaryClass = "gr-qc",
    doi = "10.1088/1361-6382/ab9bbb",
    journal = "Class. Quant. Grav.",
    volume = "37",
    number = "15",
    pages = "155003",
    year = "2020"
}

@article{East:2020hgw,
    author = "East, William E. and Ripley, Justin L.",
    title = "{Evolution of Einstein-scalar-Gauss-Bonnet gravity using a modified harmonic formulation}",
    eprint = "2011.03547",
    archivePrefix = "arXiv",
    primaryClass = "gr-qc",
    doi = "10.1103/PhysRevD.103.044040",
    journal = "Phys. Rev. D",
    volume = "103",
    number = "4",
    pages = "044040",
    year = "2021"
}

@article{Corman:2024vlk,
    author = "Corman, Maxence and East, William E.",
    title = "{Black hole-neutron star mergers in Einstein-scalar-Gauss-Bonnet gravity}",
    eprint = "2405.18496",
    archivePrefix = "arXiv",
    primaryClass = "gr-qc",
    doi = "10.1103/PhysRevD.110.084065",
    journal = "Phys. Rev. D",
    volume = "110",
    number = "8",
    pages = "084065",
    year = "2024"
}

@article{Lara:2025kzj,
    author = "Lara, Guillermo and others",
    title = "{Signatures from metastable oppositely-charged black hole binaries in scalar Gauss-Bonnet gravity}",
    eprint = "2505.14785",
    archivePrefix = "arXiv",
    primaryClass = "gr-qc",
    month = "5",
    year = "2025"
}

@article{AresteSalo:2025sxc,
    author = "Arest{\'e} Sal{\'o}, Llibert and Doneva, Daniela D. and Clough, Katy and Figueras, Pau and Yazadjiev, Stoytcho S.",
    title = "{Challenges in the nonlinear evolution of unequal mass binaries in scalar-Gauss-Bonnet gravity}",
    eprint = "2507.13046",
    archivePrefix = "arXiv",
    primaryClass = "gr-qc",
    doi = "10.1103/tr7v-jhhm",
    journal = "Phys. Rev. D",
    volume = "112",
    number = "8",
    pages = "084022",
    year = "2025"
}

@article{Corman:2025wun,
    author = "Corman, Maxence and Arest{\'e} Sal{\'o}, Llibert and Clough, Katy",
    title = "{Black hole binaries in shift-symmetric Einstein-scalar-Gauss-Bonnet gravity experience a slower merger phase}",
    eprint = "2511.19073",
    archivePrefix = "arXiv",
    primaryClass = "gr-qc",
    month = "11",
    year = "2025"
}

@article{Noakes:1983xd,
    author = "Noakes, D. R.",
    title = "{THE INITIAL VALUE FORMULATION OF HIGHER DERIVATIVE GRAVITY}",
    doi = "10.1063/1.525906",
    journal = "J. Math. Phys.",
    volume = "24",
    pages = "1846--1850",
    year = "1983"
}

@article{East:2023nsk,
    author = "East, William E. and Siemonsen, Nils",
    title = "{Instability and backreaction of massive spin-2 fields around black holes}",
    eprint = "2309.05096",
    archivePrefix = "arXiv",
    primaryClass = "gr-qc",
    doi = "10.1103/PhysRevD.108.124048",
    journal = "Phys. Rev. D",
    volume = "108",
    number = "12",
    pages = "124048",
    year = "2023"
}

@article{May:2025arz,
    author = "May, Taillte and East, William E.",
    title = "{Nonlinear evolution of the spin-2 black hole superradiant instability}",
    eprint = "2509.13488",
    archivePrefix = "arXiv",
    primaryClass = "gr-qc",
    month = "9",
    year = "2025"
}

@article{Deffayet:2025lnj,
    author = "Deffayet, C{\'e}dric and Held, Aaron and Mukohyama, Shinji and Vikman, Alexander",
    title = "{Ghostly interactions in (1+1)-dimensional classical field theory}",
    eprint = "2504.11437",
    archivePrefix = "arXiv",
    primaryClass = "hep-th",
    doi = "10.1103/bv46-rd85",
    journal = "Phys. Rev. D",
    volume = "112",
    number = "6",
    pages = "065011",
    year = "2025"
}

@article{Penrose:1969pc,
    author = "Penrose, R.",
    title = "{Gravitational collapse: The role of general relativity}",
    doi = "10.1023/A:1016578408204",
    journal = "Riv. Nuovo Cim.",
    volume = "1",
    pages = "252--276",
    year = "1969"
}

@inproceedings{Wald:1997wa,
    author = "Wald, Robert M.",
    title = "{Gravitational collapse and cosmic censorship}",
    eprint = "gr-qc/9710068",
    archivePrefix = "arXiv",
    reportNumber = "EFI-97-43",
    doi = "10.1007/978-94-017-0934-7_5",
    pages = "69--85",
    month = "10",
    year = "1997"
}

@article{Cardoso:2019rvt,
    author = "Cardoso, Vitor and Pani, Paolo",
    title = "{Testing the nature of dark compact objects: a status report}",
    eprint = "1904.05363",
    archivePrefix = "arXiv",
    primaryClass = "gr-qc",
    doi = "10.1007/s41114-019-0020-4",
    journal = "Living Rev. Rel.",
    volume = "22",
    number = "1",
    pages = "4",
    year = "2019"
}

@article{Yunes:2025xwp,
    author = "Yunes, Nicol{\'a}s and Siemens, Xavier and Yagi, Kent",
    title = "{Gravitational-wave tests of general relativity with ground-based detectors and pulsar-timing arrays}",
    doi = "10.1007/s41114-024-00054-9",
    journal = "Living Rev. Rel.",
    volume = "28",
    number = "1",
    pages = "3",
    year = "2025"
}

@article{Barack:2018yly,
    author = "Barack, Leor and others",
    title = "{Black holes, gravitational waves and fundamental physics: a roadmap}",
    eprint = "1806.05195",
    archivePrefix = "arXiv",
    primaryClass = "gr-qc",
    doi = "10.1088/1361-6382/ab0587",
    journal = "Class. Quant. Grav.",
    volume = "36",
    number = "14",
    pages = "143001",
    year = "2019"
}

@article{Ayzenberg:2023hfw,
    author = "Ayzenberg, D. and others",
    title = "{Fundamental physics opportunities with future ground-based mm/sub-mm VLBI arrays}",
    eprint = "2312.02130",
    archivePrefix = "arXiv",
    primaryClass = "astro-ph.HE",
    doi = "10.1007/s41114-025-00057-0",
    journal = "Living Rev. Rel.",
    volume = "28",
    number = "1",
    pages = "4",
    year = "2025",
    note = "[Erratum: Living Rev.Rel. 28, 7 (2025)]"
}

@article{Berti:2015itd,
    author = "Berti, Emanuele and others",
    title = "{Testing General Relativity with Present and Future Astrophysical Observations}",
    eprint = "1501.07274",
    archivePrefix = "arXiv",
    primaryClass = "gr-qc",
    doi = "10.1088/0264-9381/32/24/243001",
    journal = "Class. Quant. Grav.",
    volume = "32",
    pages = "243001",
    year = "2015"
}

@article{Figueras:2025wtx,
    author = "Figueras, Pau and Kov{\'a}cs, {\'A}ron D. and Yao, Shunhui",
    title = "{Stable non-linear evolution in regularised higher derivative effective field theories}",
    eprint = "2505.00082",
    archivePrefix = "arXiv",
    primaryClass = "hep-th",
    reportNumber = "EFI-25-4",
    doi = "10.1007/JHEP10(2025)150",
    journal = "JHEP",
    volume = "10",
    pages = "150",
    year = "2025"
}

@article{Held:2025fii,
    author = "Held, Aaron",
    title = "{Global stability of ghostly field theories: Classical scattering in $(N+1)$ dimensions}",
    eprint = "2509.18049",
    archivePrefix = "arXiv",
    primaryClass = "gr-qc",
    month = "9",
    year = "2025"
}

@book{Bronnikov:2012wsj,
    author = "Bronnikov, Kirill A. and Rubin, Sergey G.",
    title = "{Black Holes, Cosmology and Extra Dimensions}",
    doi = "10.1142/12186",
    isbn = "978-981-4374-20-0, 978-981-4440-02-8",
    publisher = "World Scientific",
    year = "2012"
}

@article{Modesto:2015ozb,
    author = "Modesto, Leonardo and Shapiro, Ilya L.",
    title = "{Superrenormalizable quantum gravity with complex ghosts}",
    eprint = "1512.07600",
    archivePrefix = "arXiv",
    primaryClass = "hep-th",
    doi = "10.1016/j.physletb.2016.02.021",
    journal = "Phys. Lett. B",
    volume = "755",
    pages = "279--284",
    year = "2016"
}

@article{Krasnikov:1987yj,
    author = "Krasnikov, N. V.",
    title = "{Nonlocal gauge theories}",
    doi = "10.1007/BF01017588",
    journal = "Theor. Math. Phys.",
    volume = "73",
    pages = "1184--1190",
    year = "1987"
}

@article{Kuzmin:1989sp,
    author = "Kuz'min, Yu. V.",
    title = "{Finite nonlocal gravity}",
    journal = "Sov. J. Nucl. Phys.",
    volume = "50",
    pages = "1011--1014",
    year = "1989"
}

@article{Tomboulis:1997gg,
    author = "Tomboulis, E. T.",
    title = "{Superrenormalizable gauge and gravitational theories}",
    eprint = "hep-th/9702146",
    archivePrefix = "arXiv",
    reportNumber = "UCLA-97-TEP-2",
    month = "2",
    year = "1997"
}

@article{Modesto:2011kw,
    author = "Modesto, Leonardo",
    title = "{Super-renormalizable quantum gravity}",
    eprint = "1107.2403",
    archivePrefix = "arXiv",
    primaryClass = "hep-th",
    doi = "10.1103/PhysRevD.86.044005",
    journal = "Phys. Rev. D",
    volume = "86",
    pages = "044005",
    year = "2012"
}

@article{Biswas:2011ar,
    author = "Biswas, Tirthabir and Gerwick, Erik and Koivisto, Tomi and Mazumdar, Anupam",
    title = "{Towards singularity and ghost free theories of gravity}",
    eprint = "1110.5249",
    archivePrefix = "arXiv",
    primaryClass = "gr-qc",
    doi = "10.1103/PhysRevLett.108.031101",
    journal = "Phys. Rev. Lett.",
    volume = "108",
    pages = "031101",
    year = "2012"
}

\end{document}